\documentstyle[12pt]{article}
\begin{document}
\vskip 2.6 truecm
\newcommand{\bb}{\begin{equation}}
\newcommand{\ee}{\end{equation}}
\newcommand{\bba}{\begin{eqnarray}}
\newcommand{\eea}{\end{eqnarray}}
\newcommand{\bban}{\begin{eqnarray*}}
\newcommand{\eean}{\end{eqnarray*}}
\newcommand{\pp}{\partial}
\newcommand{\lsim}{\buildrel <\over\sim} 
\newcommand{\gsim}{\buildrel >\over\sim}
\newcommand{\dddotV}{\buildrel{...}\over V}
\newcommand{\ddddotV}{\buildrel{....}\over V}

\begin{titlepage}
 
\title{Quantum fields interacting with colliding plane waves:
the stress-energy tensor and backreaction} \author{  {\sc  Miquel Dorca  } \\
        {\small\em Grup de F\'\i sica Te\`orica   and IFAE }\\
        {\small\em Universitat Aut\`onoma de Barcelona } \\
        {\small\em 08193 Bellaterra (Barcelona), Spain }\\
        { }\\         
        {\sc Enric Verdaguer } \\
        {\small\em Departament de F\'\i sica Fonamental }\\
        {\small\em and IFAE }\\         
        {\small\em Universitat de Barcelona } \\
        {\small\em  Av. Diagonal, 647 } \\
        {\small\em 08028 Barcelona, Spain }  }
 
\date{\null}
\maketitle
\begin{abstract}
 
Following a previous work on the quantization of a massless scalar field in a 
spacetime representing the head
on collision of two plane waves which fucus into a Killing-Cauchy horizon, we 
compute the renormalized
expectation value of the stress-energy tensor of the quantum field near that 
horizon in the physical state
which corresponds to the Minkowski vacuum before the collision of the waves. 
It is found that for minimally
coupled and conformally coupled scalar fields the respective stress-energy 
tensors are unbounded in the
horizon. The specific form of the divergences suggests that
when the semiclassical Einstein equations describing the backreaction of the 
quantum fields on the spacetime
geometry are taken into account, the horizon will acquire a curvature 
singularity. Thus the Killing-Cauchy
horizon which is known to be unstable under ``generic" classical perturbations 
is also unstable by vacuum
polarization. The calculation is done following the point splitting 
regularization technique. The dynamical
colliding wave spacetime has four quite distinct spacetime regions, namely, 
one flat region, two single plane
wave regions, and one interaction region. Exact mode solutions of the quantum 
field equation cannot be found
exactly, but the blueshift suffered by the initial modes in the plane wave and 
interaction regions makes the
use of the WKB expansion a suitable method of solution. To ensure the correct 
regularization of the
stress-energy tensor, the initial flat modes propagated into the interaction 
region must be given to a rather
high adiabatic order of approximation.
 
\end{abstract}
\end{titlepage}


\section{Introduction}
 
Exact {\it gravitational plane waves} are very simple time dependent plane 
symmetric solutions to Einstein's
equations
\cite{bel26}. Yet, as a consequence of the non linearity of General Relativity 
these solutions show some
nontrivial global features, the most conspicuous of which is the presence of a 
non-singular null Cauchy horizon
(a Killing-Cauchy horizon, in fact). This horizon may be
understood as the caustic produced by the focusing of null rays \cite{pen65}. 
The inverse of
the focusing time is a measure of the strenght of the wave, thus for an 
Einstein-Maxwell plane wave such inverse
time equals the electromagnetic energy per unit surface of the wave. This 
makes exact plane waves very
different from their linearized counterparts, which have no focusing points 
and admit a globally hyperbolic
spacetime structure. One expects that exact plane waves may be relevant for 
the study of the strong time
dependent gravitational fields that may be produced in the collision of black 
holes \cite{dea79,fer80} or to
represent travelling waves on strongly gravitating cosmic strings 
\cite{gar89-90}. In recent years these
waves have been used in classical general relativity to test some conjectures 
on the stability of Cauchy
horizons
\cite{ori92,yur93}, and in string theory to test classical and quantum string 
behaviour in strong
gravitational fields \cite{veg84-90,veg91,jof94}. Their interest also stems 
from the fact that plane waves are a
subclass of exact classical solutions to string theory 
\cite{ama84-88,hor90,tse93fes94rus95}. 
 
When plane waves
are coupled to quantum fields the effects are rather trivial since they 
produce neither vacuum polarization
nor the spontaneous creation of particles, in that sense these waves behave 
very much as electromagnetic or
Yang-Mills plane waves in flat spacetime \cite{des75,gib75}. Still the 
classical focusing of geodesics has a
quantum counterpart: when quantum particles are present the quantum field 
stress-energy tensor between
scattering states is unbounded at the Cauchy horizon, i.e. where classical 
test particles focus after colliding
with the plane wave
\cite{gar91}. This suggests that the Cauchy horizon of plane waves may be 
unstable under the presence of quantum
particles. The classical instability of the null Cauchy horizons of plane 
waves is manifest when non-linear
plane symmetric gravitational radiation collides with the background wave, 
i.e. when {\it two plane
waves collide}. In this case the focusing effect of each wave distorts the 
causal structure of the spacetime
near the previous null horizons and either a spacelike curvature singularity 
or a new regular Killing-Cauchy
horizon is formed. However, it is generally believed that the Killing-Cauchy 
horizons of the colliding plane
wave spacetimes are unstable in the sense that ``generic" perturbations will 
transform them into spacelike
curvature singularities. In fact, this has been proved under general plane 
symmetric perturbations
\cite{yur88-89}. Also exact colliding plane wave solutions with classical 
fields are known that have
spacelike curvature singularities and which reduce, in the vacuum limit, to 
colliding plane wave solutions
with a regular Killing-Cauchy horizon
\cite{cha87}.
 
{\it Colliding plane wave spacetimes} are some of the simplest dynamical 
spacetimes and, as such, they have been
used as a test bed for some problems in classical general relativity such as 
the, just mentioned, stability of
the Killing-Cauchy horizons, or the cosmic censorship hypothesis \cite{yur93}. 
Note that in a colliding plane
wave spacetime with a Killing-Cauchy horizon inequivalent extensions can be 
made through the horizon, and this
implies a breakdown of predictability since the geometry beyond the horizon is 
not uniquely determined by the
initial data posed by the incoming colliding plane waves. The singularities in 
colliding wave spacetimes are
also different from the more familiar cosmological and black hole 
singularities which are originated by the
collapse of matter since they result from the non linear effects of pure 
gravity. The type of singularities
also differs in the sense that these are all encompassing, i.e. all timelike 
and null geodesics will hit the
singularity in the future.
 
In a previous paper \cite{dor93} which we shall refer to as paper I, we made a 
first step in the study of the
interaction of massless scalar quantum fields with a gravitational background 
which represents the head on
collision of two linearly polarized shock waves followed by trailling 
gravitational radiation which focus into
a Killing-Cauchy horizon \cite{fer87cha86,hay89}. The spacetime is divided 
into four distinct regions: a
flat space region (which represents the initial flat region before the waves 
collide), two single plane wave
regions (the plane waves before the collision) and the interaction region 
which is
bounded by the previous three regions and a regular Killing-Cauchy horizon. 
The interaction region is locally
isometric to a region inside the event horizon of a Schwarzshild black hole 
with the Killing-Cauchy horizon
corresponding to the event horizon.
The presence of the Killing horizon made possible the definition of a natural 
preferred ``out" vacuum state
\cite{kay91} and it was found that the initial flat vacuum state contains a 
spectrum of ``out" particles. In the
long-wave lenght limit the spectrum is consistent with a thermal spectrum at a 
temperature which is inversely
proportional to the focusing time of the plane waves. Of course, the 
definition of such ``out" vacuum is not
possible when we have a curvature singularity (i.e. in the ``generic" case) 
instead of an horizon, whereas a
physically meaningful ``in" vacuum may be defined in all colliding plane wave 
spacetimes.
 
In this paper we compute the expectation value of the stress-energy tensor of 
the quantum field
near the horizon in the initial
flat space vacuum. The expectation value of the stress-energy tensor is  the 
relevant
observable in the quantization of a field in a gravitational background since 
it is the source of gravity (to
be written to the right hand side in the semiclassical modification of 
Einstein's equations). These
semiclassical equations are interpreted as dynamical equations for both the 
quantum field and the gravitational
field and determine the backreaction of the quantum field on the spacetime 
geometry. We find, not
surprisingly, that the stress-energy tensor is unbounded at the horizon. The 
specific form of this divergence
suggests that when the backreaction is taken into account the horizon will 
become a spacetime singularity,
i.e. the Killing-Cauchy horizon is unstable under vacuum polarization. Note 
that this is a non perturbative
effect, it is the result of the nonlinearity of gravity, since gravitational 
waves in the linear approximation
do not polarize the vacuum. In fact the vacuum stress-energy tensor of a 
quantum field in a weakly inhomogeneous
background was computed by Horowitz \cite{hor80}, and it is easy to see that 
such tensor can be written in
terms of the linearized Einstein tensor only \cite{cam94}, which vanishes for 
gravitational waves.
 
The non perturbative evaluation of the expectation value of the stress-energy 
tensor of a quantum field in a
dynamically evolving spacetime is generally a difficult task. So far the most 
relevant calculation in this
respect is, probably, that of the expectation value of the stress-energy 
tensor in the Unruh vacuum state of a
blak hole near its horizon \cite{can80sci81,fro85tho86}. Note that although 
such calculation is done in the
extended Schwarzschild spacetime rather than in the dynamical spacetime 
describing the gravitational collapse,
the Unruh vacuum describes in the extended spacetime the vacuum state with 
respect to modes which are incoming
from infinity in the collapsing spacetime.
 
Even when the exact modes of the quantum field equation are known it may not 
be possible to perform the mode
sums in order to get the quantum field two point function or, more precisely, 
the Hadamard function, which is
the key ingredient in the evaluation of the stress-energy tensor. In our 
colliding wave spacetime we do not
even know the exact solution of the modes in the interaction region (a similar 
situation is produced in the
Schwarzschild case). Fortunatelly the geometry of the colliding spacetime is 
such that the initial modes which
come from the flat region are strongly blueshifted in their frequency in  the 
interaction region near the
horizon. This makes the use of the  geometrical optics approximation and, more 
generally, of a systematic WKB
expansion of the modes in the interaction region a very useful tool.
Here we should mention that the blueshift of the initial modes in the 
interaction region is not exclussive
of the particular colliding wave spacetime that we take, it is a general 
feature in colliding wave spacetimes
and it is due to the focusing produced by the initial plane waves. This fact 
has also been exploited,
for instance, to compute the production of particles in the so called 
Bell-Szekeres colliding spacetime
\cite{fei95}.
 
The plan of the paper is the following. In section 2 the geometry of the 
colliding plane wave spacetime is
briefly reviewed. In section 3 the mode
solutions of the scalar field equation are given for the four different 
regions of the spacetime, it is only
in the interaction region that exact solutions for these modes cannot be 
found. In section 4 the geometrical
optics approximation is used to relate the parameters of the modes in the 
initial flat region with the
parameters of the modes propagated to the interaction region, in this way a 
physical meaning for these
parameters is found. In section 5 the initial flat spacetime modes which
are propagated in the interaction region are given in terms of a WKB expansion 
up to the adiabatic order four.
This is the order which will be required for the regularization of the 
Hadamard function. Then in section 6
the point splitting technique is reviewed for the computational purposes of 
this paper and 
in the short section 7 the renormalized
stress-energy tensor is computed in the single plane wave regions, this is not 
necessary since we know from the
literature
\cite{gib75,vil77} that it vanishes, but this calculation is a simple and 
illustrative application of the point
splitting regularization technique.
In section 8, which
is the core of the paper, this technique is used to regularize the Hadamard 
function and
calculate its value by a mode sum near the Killing-Cauchy horizon. Finally, in 
section 9 the expectation
value of the stress-energy tensor near the horizon is calculated. A summary 
and some consequences of
our results, such as the backreaction problem, the quantum instability of the 
Killing-Cauchy horizon and the
generality of these results are discussed in section 10. In order to keep the 
main body of the paper reasonably
clear, many of the technical details of the calculations, as well as a short 
review on the algebra of
bitensors, have been left to the Appendices.

\section{Geometry of the colliding plane wave spacetime}
 
Here we recall the main geometric properties of a spacetime describing
the head on collision of two linearly polarized gravitational plane waves
propagating in the $z$-direction and forming a regular Killing-Cauchy horizon; 
further details can be found in
paper I. This spacetime has four regions (see Fig. 1): a flat region (or 
region IV) at the past,
before the arrival of the waves, two plane wave regions (regions II and III) 
and an interaction region (region
I) where the waves collide and interact nonlinearly. The geometry of these 
regions is described by the
following four metrics, which are solutions of the Einstein's field equations 
in vacuum and are written in
coordinates adapted to the two commuting Killing vector fields ${\partial}_x$ 
and ${\partial}_y$,
  
$$ds_{\rm
I}^2=4{L_1L_2}\,{\left[{1+\sin\left({u+v}\right)}\right]}^{2}dudv-{1-\sin%
\left({u+v}\right) \over
1+\sin\left({u+v}\right)}d{x}^{2}-$$ \bb
-{\left[{1+\sin\left({u+v}\right)}\right]}^{2}{{\cos}^{2}\left({u-v}%
\right)}d{y}^{2}{}^{} , \label{eq:MI}\ee
 
\bb
{ds^2}_{\rm II}=4{L_1L_2}\,{\left({1+\sin u}\right)}^{2}dudv-{1-\sin u
\over 1+\sin u}d{x}^{2}
-{\left({1+\sin u}\right)}^{2}{{\cos}^{2} u}d{y}^{2}{}^{},
\label{eq:MII}\ee
 
\bb
{ds^2}_{\rm III}=4{L_1L_2}\,{\left({1+\sin v}\right)}^{2}dudv-{1-\sin v
\over
1+\sin v}d{x}^{2}
-{\left({1+\sin v}\right)}^{2}{{\cos}^{2} v}d{y}^{2}{}^{},
\label{eq:MIII}\ee
 
\bb{ds^2}_{\rm IV}=4{L_1L_2}\,dudv-d{x}^{2}-d{y}^{2}, \label{eq:MIV}\ee
where $u$ and $v$ are two dimensionless null coordinates ($v+u$ is a time 
coordinate and $v-u$ a space
coordinate) and $L_1$, $L_2$ are two arbitrary positive lenght parameters, 
which represent the 
focusing time (i.e. the inverse of the strength) of the plane waves. The 
boundaries of these four regions are:
$\{u=0$,
$v\leq 0\}$ between regions IV and II, $\{v=0$, $u\leq 0\}$ between regions IV 
and III, $\{v=0$, $0\leq u<{\pi}/2\}$
between regions II and I and $\{u=0$, $0\leq v<{\pi}/2\}$ between regions III 
and I. At these boundaries the
matching of the metrics is such that the Ricci tensor vanishes, $R_{\mu\nu}=0$ 
(i.e. we have a vacuum solution
in the entire spacetime).
 
At the boundaries $u={\pi}/2$ and $v={\pi}/2$ on regions II and III, 
respectively, the determinants of the respective
metrics vanish, these are the points of focusing of the plane waves and are 
coordinate singularities. The
causal structure of these regions is best described with the
use of appropriate coordinates (harmonic coordinates). In these new 
coordinates the boundaries $u={\pi}/2$ and
$v={\pi}/2$ are seen to be spacetime lines.
 
Region I
is locally isometric to a  region of the
Schwarzschild metric bounded by the event horizon. This is easily seen with 
the coordinate
transformation,
 
$$t=x,\;\;\;\; r=M\left[{1+\sin (u+v)}\right],\;\;\;\;\varphi =1+y/M,\;\;\;\; 
\theta = {\pi}/2-(u-v)$$
where we have defined $M=\sqrt{L_1 L_2}$. Then the metric (\ref{eq:MI}) becomes
 
$$ ds^2=\left({{2M \over r}-1}\right)^{-1}d{r}^{2}-\left({{2M
\over r}-1}\right)d{t}^{2}-{r}^{2}\left({d{\theta }^{2}+{{\sin}^{2}\theta
}d{\varphi }^{2}}\right), $$
which is the interior of the Schwarzshild black hole. The surface $u+v={\pi}/2$
corresponds to the black hole event horizon. The boundary $v=0$ corresponds to
$r=M(1+\cos\theta)$ and $u=0$ corresponds to $r=M(1-\cos\theta)$, these are the
boundaries of the plane waves. These boundaries join at
$r=M$ (spacetime point of the collision) and also at the surface $u+v={\pi}/2$ 
at $\theta
=0$ and $\theta ={\pi}$. This region of the black hole interior is outside the
singularity $r=0$ and thus the interaction region has no curvature
sin\-gu\-la\-ri\-ties. The above local isometry is not global however, the
coordinates $\theta$ and $\phi$ are cyclic in the black hole case but in the
plane wave case, $-\infty <y<\infty$ and $-\infty <v-u<\infty$.
 
 
As in the the Schwarzschild case it is
convenient to introduce a set of Kruskal-Szekeres like coordinates to describe
the interaction region, because the $(u,\, v,\, x,\, y)$ coordinates become 
singular at the horizon.
Since the  Klein-Gordon equation can be separated in the new coordinates, 
these will play an important role
in the quantization. 
First we introduce dimensionless time and space
coordinates $(\xi ,\eta)$ 
\bb \xi =u+v,\;\; \eta =v-u
,\label{eq:XIETA}\ee 
with the range
$0\leq\xi <{\pi}/2,\;\; -{\pi}/2\leq\eta <{\pi}/2$. Then we introduce a new 
time coordinate ${\xi}^*$
related to the dimensionless time coordinate $\xi$ by 
 
\bb {\xi
}^{{}^*}=2M\ln\left({{1+\sin\xi  \over 2\,{{\cos}^2\xi 
}}}\right)-M\left({\sin\xi
-1}\right) ,\label{eq:XI*}\ee
and a new set of null coordinates
${\tilde U} ={\xi}^*-x$,
${\tilde V}={\xi}^*+x$.
Note that the transversal coordinate $x$, which behaves badly at the horizon, 
appears in this coordinate
transformation. Finally, we define, 
 
\bb U'=-2M{\rm exp}\left({-{{\tilde
U}/4M}}\right)\leq 0,\;\; V'=-2M{\rm exp}\left({-{{\tilde V}/4M}}\right)\leq 
0,\label{eq:U'V'}\ee 
and the metric in the interaction region (\ref{eq:MI}) can be written as,
$$  d{s}^{2}_I={{2\,{\rm e}^{(1-\sin\xi )/2} \over
\left({1+\sin\xi }\right)}}dU'dV'-{M}^{2}{\left({1+\sin\xi
}\right)}^{2}d{\eta }^{2}-
{\left({1+\sin\xi }\right)}^{2}{\cos^{2}\eta }d{y}^{2} ,\label{eq:HM}$$
with
 
$$ U'V'=8{M}^{2}{{{\cos}^{2}\xi } \over 1+\sin\xi }{\rm exp}\left({{\sin\xi -1
\over 2}}\right) ,\;\;\;\; 
{ U' \over V'}={\rm exp}\left({{x \over 2M}}\right).$$
The curves $\xi ={\rm const.}$ and $x={\rm const.}$ are, respectively, 
hyperbolae
and straight lines through the origin of coordinates $(U'=V'=0)$.
The Schwarzshild horizon (which is a Killing-Cauchy horizon for our spacetime) 
corresponds to the limit of the
hyperbolae when $\xi\rightarrow {\pi}/2$ i.e. $ V'=0$ or $U'=0$. Notice that the
problem with the transversal coordinate $x$ at the horizon is that all the lines
$x={\rm const.}$ go through the origin of the $(U',V')$ coordinates, so that 
all the
range of $x$ collapses into the point $V'=U'=0$, whereas the lines $U'=0$ and
$V'=0$ represent $x=-\infty$ and $x=\infty$ respectively. One should  recall
that we have not represented the coordinate $x$ in Fig. 1 where only the 
$(u,v)$ coordinates are shown, $x$
is a transversal coordinate perpendicular to the propagation of the waves.
 
To understand the global geometry of the spacetime a tridimensional picture 
like Fig. 3 of paper I, is helpful.
It represents the boundary surfaces between the different regions in terms of 
the appropriate
nonsingular coordinates adapted to each region. The spacetime lines ${\cal L}$ 
$(u={\pi}/2)$ and 
${\cal L}'$ $(v={\pi}/2)$ are
identified with points ${\cal P}$ and ${\cal P}'$ respectively, these points 
are known as {\it folding
singularities}, and are avoided by all null geodesics except for a set of null 
measure. There is no spacetime
beyond the lines ${\cal L}$ and ${\cal L}'$ (see Fig. 1).

\section{Mode solutions of the field equation}
 
In order to quantize a field in our background spacetime and to compute the 
field two-point function which
will be needed later, we need the mode solutions of the field equations. Here 
we consider a massless scalar
field
$\phi$, the Klein-Gordon equation is
 
\bb \Box\phi =0, \label{eq:KG'}\ee
where $\Box\phi
=(-g)^{-1/2}[(-g)^{1/2}g^{\mu\nu}\phi _{,\nu}]_{,\mu}$, and $g$ is the 
determinant of the metric
$g_{\mu\nu}$. Note that since the curvature scalar $R$ is zero in our case, 
the above equation is valid whatever
coupling to the curvature we may consider. The mode solutions of these 
equations are different in the four
spacetime regions we have. In paper I details of the equations in the 
different regions and their solution by
separation of variables were given. Here we summarize the main results and 
introduce some new ones which will
be of use later.
 
Our vacuum state is the ``in" vacuum, this is the physically unambiguous 
vacuum defined in the
flat region IV before the plane waves arrive. In this region a complete set of 
positive frequency modes
with respect to the timelike Killing vector $\partial _{u'+v'}$ is given by:
 
\bb {u}_{k}^{{\rm in}}(u,v,x,y)={1\over \sqrt{2{{k}}_- {(2{\pi})}^3 }
}\,{\rm e}^{-i2L_2k_- v -i2L_1k_+ u +ik_x x+ik_y y} ,\label{eq:m-iv}\ee
where $u$ and $v$ are the two dimensionless null coordinates related to a 
physical null coordinates $u'$ and
$v'$ by $u'=2L_1u$, $v'=2L_2v$.
The labels $k_x$, $k_y$ and $k_-$ are independent separation constants for 
equation
(\ref{eq:KG'}) and the label $k_+$  is determined by the relation,
 
\bb 4k_+k_-=k_x^2+k_y^2.\label{eq:r-iv}\ee 
It was shown in paper I that these modes are well normalized on the
hypersurface $\{u=0,\, v\leq 0\}\cup\{v=0,\, u\leq 0\}$.
The labels $k_x$ and $k_-$ are continuous but $k_y$ is discrete if we take a 
cyclic spacetime
in the $y$-direction (this is convenient but not necessary), then we identify 
$k_y$ with $mM^{-1}$
where
$m$ is an integer.  
 
These ``in" modes, ${u}_{k}^{{\rm in}}$, which define our vacuum state, have 
to be propagated through the
spacetime. Once we know the ``in" modes in a certain region we use their 
expression on the boundaries with
the next region as boundary conditions for the ``in" modes in this next 
region. These will be given 
in terms of a complete set of
solutions  of the field equation in such region. In the plane wave
regions II and III the ``in" modes are easy to find:
 
\bb u_k^{\rm in}(x)={1\over\sqrt{2k_-(2\pi )^3}}{\rm e}^{ik_x x+ik_y 
y}\left\{\begin{array}{l}
\displaystyle{1\over\cos u }{\rm e}^{-i2L_2k_- v-iA(u)/(2L_2k_-)},\;\;\;{\rm 
Region \ II}\\
\\
\displaystyle{1\over\cos v }{\rm e}^{-i2L_1k_+ u-iA(v)/(2L_1k_+)},\;\;\;{\rm 
Region \ III}
\end{array}\right.      \label{eq:m-ii}\ee 
where $A(x)=k_1^2\, f(x)+k_2^2\, g(x)$ and where the
two fuctions $f(x)$ and $g(x)$ are 
 
\bb f(x)={{{\left({1+\sin x}\right)}^{2} \over
2\,\cos x}\left({9-\sin x}\right)+{15 \over 2}\cos x-{15 \over
2}\, x-12},\;\;\;
g(x)=\tan x.\label{eq:fg}\ee
We use for simplicity the dimensionless labels $k_1=Mk_x$ and $k_2=Mk_y$ where 
$k_x$, $k_y$ and
$k_-$ have the same meaning as in the flat region IV because these expressions 
for the solutions of the
Klein-Gordon equation in region II and III match smoothly (i.e in a continuous 
and differentiable way) with
the respective solutions (\ref{eq:m-iv}) on the boundary between regions II 
and IV, i.e
$\{u=0$, $v\leq 0\}$ and the boundary between regions III and IV, i.e $\{v=0$, 
$u\leq 0\}$.
 
In the interaction region (region I), however, this kind of mode propagation 
is not that simple.
First, we note that equation (\ref{eq:KG'}) in this
region can be separated as  
 
\bb \phi (\xi ,\eta
,x,y)=e^{ik_x x+ik_y y}\,\psi _{\alpha k_x}(\xi)\,\varphi _{\alpha k_y}(\eta), 
 \ee 
where coordinates $\xi$, $\eta$ are related to the usual null coordinates
$u$, $v$, by (\ref{eq:XIETA}) and the
differential equations for $\psi _{\alpha
k_x}(\xi)$ and $\varphi _{\alpha k_y}(\eta)$ are (dropping the indices)
 
\bb {\psi}_{,\xi \xi}-(\tan\xi
){\psi}_{,\xi }+\left( \alpha +k_1^2{{(1+\sin\xi )}^4 \over
{\cos}^2 \xi}\right)\psi = 0, \label{eq:e1}\ee
 
\bb {\varphi }_{,\eta \eta}-(\tan\eta ){\varphi}_{,\eta }+\left( \alpha
-k_2^2{1 \over {\cos}^2 \eta}\right)\varphi = 0  , \label{eq:e2} \ee
where $\alpha$ is a dimensionless separation constant and $k_x$, $k_y$ (or 
$k_1=Mk_x$, $k_2=Mk_y$) are the
same labels as in regions IV, II or III. 
 
The solutions of (\ref{eq:e2}) can be expressed in terms of associated 
Legendre functions and in the cyclic
case 
they can be written in terms of spherical harmonics, $Y^m_l(y/M,\pi /2-\eta)$, 
with discrete labels related to
$\alpha$ and $k_2$  simply by $\alpha =l(l+1)$ and $k_2=m$. The solutions of 
(\ref{eq:e1}), however, cannot be
written in terms of known functions. If we use the variable $\xi ^*$ instead 
of $\xi$ defined in 
(\ref{eq:XI*}), i.e. $d\xi =\cos\xi\, M^{-1}(1+\sin\xi)^{-2}\, d\xi ^*$, and 
introduce a new function $\gamma$
defined by
$2\,\gamma =(1+\sin\xi)\,\psi$, equation (\ref{eq:e1}) becomes
 
\bb \gamma _{,\xi ^*\xi ^*}+\omega _1^2\gamma =0,   \label{eq:e1*}\ee
where
 
\bb \omega ^2_1(\xi)\equiv M^{-2}\left[k_1^2+{\cos ^2\xi\over 
(1+\sin\xi)^4}\left(l(l+1)+{2\over
1+\sin\xi}\right)\right]
\equiv M^{-2}\left[k_1^2+V_1(\xi )\right]. \label{eq:omegaxi}\ee
Note that a new function $V_1(\xi )$ has been defined. Equation (\ref{eq:e1*}) 
can be solved by a WKB method
and the solutions
$\gamma (\xi ^*)$ can be expanded up to any adiabatic order as
 
\bb \gamma (\xi ^*) ={1\over\sqrt{W_1(\xi)}}\, {\rm e}^{\pm i\int 
_0^{\xi}W_1(\xi ') d\xi
'^*},\label{eq:egam}\ee 
where $W_1(\xi)$ is given, up to adiabatic order 4 (this is the order we need)
\cite{chak73}, by
 
\bb W_1(\xi )=\omega _1 +{A_2\over\omega _1^3} +{B_2\over\omega _1^5} 
+{A_4\over\omega _1^5}
 +{B_4\over\omega _1^7} +{C_4\over\omega _1^9} +{D_4\over\omega 
_1^{11}},\label{eq:Wxi}\ee
where,
 
\bb A_2=-{1\over 8}\,{{\ddot V}_1\over M^2},\;\;\;\; B_2={5\over 35}\,{{\dot 
V}^2_1\over M^4},\label{eq:An}\ee
 
$$ A_4={1\over 32}\,{{\ddddotV}_1\over M^2},\;\;\;\; B_4=-{1\over 
128}\,{28\,{\dot V}_1\,{\dddotV}_1+19\,{\ddot
V}^2_1\over M^4},\;\;\;\; C_4={221\over 258}\,{{\dot V}^2_1\,{\ddot V}_1\over 
M^6},\;\;\;\; D_4=-{1105\over
2048}\,{{\dot V}^4_1\over M^8},$$ 
an overdot means derivative with respect to $\xi ^*$, and 
$A_n$, $B_n$, ... denote the $n$ adiabatic terms in
$W_1(\xi)$. From the two exponential factor signs in the WKB solution 
(\ref{eq:egam}),
we choose the negative sign for consistency with the boundary conditions 
imposed by (\ref{eq:m-ii}), as we will
see later. This gives
positive frequency solutions with respect to the timelike vector 
$d/d\xi ^*$ on the horizon since these solutions are asymptotically 
proportional to 
${\rm exp}({-i|k_x|\xi ^*})$. A complete set of solutions of the Klein-Gordon 
equation in the interaction
region is thus given, in the cyclic case, by
 
\bb \phi _k (x)={2\, C\over 1+\sin\xi}{1\over\sqrt{W_1(\xi)}}\,
{\rm e}^{ik_x x-i\int _0^{\xi}W_1(\xi ') d\xi '^*}\, Y_l^m(y/M,\pi 
/2-\eta),\label{eq:phi}\ee
where $C$ is a normalization constant which can be easily calculated imposing 
that these solutions are well
normalized on the horizon ($\xi =\pi /2$). We will find its value in a next 
section.
 
Finally, the ``in" modes in the interaction region, with the boundary 
conditions imposed by (\ref{eq:m-ii}),
can be written as, 
 
\bb u_k^{\rm in} (\xi ,\eta ,x,y) =\sum _l C_l\, \phi _k (\xi ,\eta ,x,y)
,\label{eq:lin}   \ee  
where the coefficients $C_l$ depend on $l$ and on the separation constants
used to label the modes in region II, i.e $k_x$, $k_y$, $k_-$. We devote the 
next two sections to find
an explicit expression for (\ref{eq:lin}).

\section{Geometrical optics}
 
The solutions of the Klein Gordon equation in the flat and single plane wave 
regions,
(\ref{eq:m-iv}) and
(\ref{eq:m-ii}) respectively, can be understood in terms of geometrical optics 
because the geometrical
optics approximation is exact in these regions, i.e. the orthogonal  curves to 
the surfaces of constant phase
(the rays) follow exactly null geodesics. This will allow to
``localize" the Klein-Gordon solutions in the spacetime and to determine which 
solutions will be
relevant in later calculations. The geometrical optics also gives information 
on the  meaning of the parameters
which label the ``in" modes (\ref{eq:m-ii}), i.e. $k_{\pm}$, $k_x$, $k_y$, and 
the parameters
$k_x$, $m$,
$l$, which label the modes (\ref{eq:phi}) in the interaction region, and on 
their connections.
 
In the flat region the normalized ``in" modes (\ref{eq:m-iv}) are purely flat 
modes which have
positive frequency with respect to the timelike Killing vector $\partial 
_t=\partial _{u'+v'}$. The energy of
these modes,
$E_{\rm IV}$, may be defined as the eigenvalue of $\partial _t$, $\partial _t\,
u_k^{\rm in(IV)}=-iE_{\rm IV}u_k^{\rm in(IV)}$. In terms of the null momenta 
$k_{\pm}$ it is,
 
\bb E_{\rm IV}=k_++k_-. \label{eq:E4}\ee
We can now proceed in the same way in the single plane wave regions. In region 
II, the normalized ``in" modes
take the form (\ref{eq:m-ii}), but now, instead of a timelike vector, 
$\partial _t$, we have a null
Killing vector $\partial _{v'}$, and (\ref{eq:m-ii}) are its eigenfunctions 
with positive eigenvalue $k_-$,
i.e.
$\partial _{v'}\, u_k^{\rm in(II)}=-ik_-u_k^{\rm in(II)}$. Similarly, in 
region III, $k_+$
are eigenvalues of $\partial _{u'}$. In this case, however,
these eingenvalues cannot be directly interpreted as energies. The mode
solutions in the plane wave regions II and III have the form
$C\, {\rm e}^{iS}$, where $S={\rm constant}$ define the surfaces of constant 
phase of the
modes and it satisfies exactly the null condition 
$g^{\mu\nu}S_{,\mu}S_{,\nu}=0$. Thus, we might define the
energy of these modes as the variation of $S$ in the direction of the vector
$-\partial _t$, i.e.
$E_{\rm II/III}=-{\partial _t}S_{\rm II/III}.$
From (\ref{eq:m-ii}), we identify
 
$$S_{II}=-2L_2k_-v-{f(u)k_1^2+g(u)k_2^2\over 2L_2k_-}+k_x x+k_y y,             
$$
and $S_{III}$, by a similar expression. The tangent vector fields to the 
congruencies of null geodesics, 
expresed in $(u,\, v,\, x,\, y)$ components are, in
region II,
 
$$V^{\mu}_{II}={-1\over 2M^2(1+\sin u)^2}\,\left(2L_2k_-,\, 
{{\dot f}(u)k_1^2+{\dot g}(u)k_2^2\over 2L_2k_-},\, 2Mk_1{\dot f}(u),\,
2Mk_2{\dot g}(u)\right),$$
and a similar expression for $V^{\mu}_{III}$ in region III.
 
By integration we obtain a congruence
of null geodesics parametrized by their momenta $k_\pm$, $k_x$, $k_y$ and 
their initial positions
$v_0$, $x_0$, $y_0$ at $u=0$ in region II and $u_0$, $x_0$, $y_0$ at $v=0$ in 
region III. Each
congruence at fixed values of the momenta represents the set of rays 
orthogonal to the surfaces
$S_{\rm II/III}= {\rm constant}$, so that these rays ``localize" in the 
spacetime the mode labeled by
$k_\pm$, $k_x$, $k_y$. A simple inspection of the expression
for the tangent vector fields above shows that in the plane wave region II the 
rays for each
mode are peaked near the line ${\cal L}\, (u=\pi /2)$ and reach the 
interaction region at
points close to the folding singularity ${\cal P}\,(u=\pi /2,\, v=0)$ if the 
momenta satify the
conditions,
 
\bb 2L_2k_-\gg 1,   \;\;\;\; {16k_1^2+k_2^2}\ll 2L_2k_-,\label{eq:kkk-}\ee
where the second condition comes from the property,
 
$${{\dot f}(u)k_1^2+{\dot g}(u)k_2^2}\leq {(16k_1^2+k_2^2)}\,\cos ^{-2}u,$$ 
of the functions $f$ and $g$, defined in (\ref{eq:fg}). Similar conditions 
hold for modes in the plane
wave region III, by the simple substitution of $2L_2k_-$ by $2L_1k_+$, i.e.
the rays for each
mode are peaked near the coordinate singularity line ${\cal L}'$ ($v=\pi /2$) 
and reach the interaction region
at points close to the folding singularity ${\cal P}'$ $(u=0,\, v=\pi /2)$ if
 
\bb 2L_1k_+\gg 1,   \;\;\;\; {16k_1^2+k_2^2}\ll 2L_1k_+.\label{eq:kkk+}\ee
Due to the relation (\ref{eq:r-iv}), the two conditions (\ref{eq:kkk-}) and 
(\ref{eq:kkk+}) are
mutually exclusive. These geometrical properties of the ``in" modes
indicate that any calculation involving ``in" modes in regions close to the 
folding singularity
${\cal P}$, requires only modes in region II with momenta satisfying 
(\ref{eq:kkk-}) and no
modes in region III.
Equivalently any calculations in regions close to the folding singularity 
${\cal P}'$,
requires only modes in region III with momenta satisfying (\ref{eq:kkk+}) and 
no modes in region II.
Note that the larger is the energy $E_{IV}$ of the ``in" modes in the flat 
region, the closer their rays get to
the lines ${\cal L}$ or ${\cal L}'$, and this means that the ``in" modes are 
{\it blueshifted} towards
these lines and consequently towards the horizon.
In fact, this is a general 
property of plane waves \cite{gar91} and it will become very important in the 
evaluation of the vacuum
expectation value of the stress-energy tensor when summations over ``in" modes 
are performed.
 
Going back to the energy of the ``in" modes in the plane wave region II we 
have, according to the previous
definition,
$E_{II}=-{\partial _t} S_{{II}}=k_-+({\dot f}(u)k_x^2+{\dot g}(u)k_y^2)\, 
(4k_-)^{-1}$, which is
not constant, because $\partial _t$ is not a
Killing vector, and cannot be interpreted properly as the energy of the mode. 
However, 
close to the origin
$u=v=0$, i.e. just where the plain waves collide,
$E_{\rm II}$ coincides with $E_{\rm IV}$, since
 
$$\left. E_{II}\right|_{u,v\sim 0}=k_-+{k_x^2+k_y^2\over 4k_-} +O(u^2)=
k_-+k_+ +O(u^2)\sim E_{IV},  $$
and close to the line $\cal L$ we obtain, in powers of $\cos u$,
 
$$E_{II}=k_-+{16k_x^2+k_y^2\over 4k_-}\,{\cos ^{-2}u} +O(\cos ^0u).  $$
However, if we recall that for the ``in" modes close to the line $\cal L$ in 
region II the inequalities
(\ref{eq:kkk-}) are satisfied, then there is a large interval of the variable 
$u$ for which,
 
\bb {16k_1^2+k_2^2\over 2L_2k_-}\,{\cos ^{-2}u}\ll 1.  \label{eq:kkkcosu}\ee
In this interval $E_{\rm II}\simeq k_-\simeq E_{IV}$,
which is the eigenvalue of the null Killing vector $\partial _{v'}$, thus it 
can be interpreted
unambiguously as the energy of the ``in" modes in region II. 
Similarly for ``in" modes close to de line ${\cal L}'$ in region III there is
a large interval of the variable $v$ for which,
 
\bb {16k_1^2+k_2^2\over 2L_1k_+}\,{\cos ^{-2}v}\ll 1.   \label{eq:kkkcosv}\ee
and their energy 
is unambiguously defined by
$E_{\rm III}\simeq k_+\simeq E_{IV}$. We will see in the next section that 
these intervals, (\ref{eq:kkkcosu})
and (\ref{eq:kkkcosv}), play an important role.

In region I, the mode solutions of the Klein-Gordon equation are given by 
(\ref{eq:phi}).
Although the geometrical optics approximation is not exact in this region we 
can still make good use of
it, since this approximation is compatible with the WKB expansion of equation 
(\ref{eq:e1*}), which depends
on the dimensionless timelike coordinate $\xi =u+v$ (or $\xi ^*$). The energy 
of these modes can be defined
by the variation of the surfaces of constant phase,
$S=-\int _0^{\xi}W_1(\xi')d\xi '^*+ik_x x+ik_y y$,
along the
timelike vector $M^{-1}\,\partial _\xi$, i.e,
$E_{I}=-M^{-1}\,{\partial _\xi}S_{\pm}$.
If we consider the range of values $\rho\equiv |k_1|L(\cos\xi )^{-1}\ll 1$ in 
analogy with
(\ref{eq:kkkcosu}) and (\ref{eq:kkkcosv}), 
$E_{\rm I}\sim M^{-1}{\left[1+O(\rho ^2)\right]\, L^{-1}}$.
In other words, the parameter $M^{-1}L^{-1}=M^{-1}[l(l+1)]^{1/2}$, that labels 
the mode solutions of the
Klein-Gordon equation in the interaction region, takes the role of an energy.
Recall that this energy is positive just because we have choosen the minus 
sign in the WKB solution 
(\ref{eq:egam}).
This suggests that we can take the
value for the ``in" modes in the interaction region close to the horizon, 
$u_k^{\rm in}$, as a single function
$\phi _k$ (\ref{eq:phi}) instead of the linear superposition (\ref{eq:lin}), 
provided that the energy of
$\phi _k$, i.e
$E_I$ equals to $E_{\rm II}$ and to $E_{\rm III}$, at the boundaries
of regions II and III with region I, respectively. In other words we can take 
$u_k^{\rm in}\sim\phi _k$ where
$\phi _k$ is labeled by a parameter $L$ satisfying,
 
\bb E_{I}\sim{M^{-1}L^{-1}}\propto k_-+k_+.   \label{eq:Lk+-}\ee
In the next section we will see that this identification is really possible 
and we will find the
proportionality constant. Note that the ``in" modes which are relevant at the 
boundary between region II
and region I satisfy $2L_2k_-\gg 1$ and $2L_1k_+\sim 0$, so that $E_{\rm 
I}\sim E_{\rm II}$ and
$L\propto L_2k_-$, and the
modes which are relevant at the boundary between region III and region I 
satisfy $2L_1k_+\gg 1$
and $2L_2k_-\sim 0$, then $E_{\rm I}\sim E_{\rm III}$ and $L\propto L_1k_+$.

\section{Adiabatic expansion of the ``in" modes}
 
We seek a solution of the
Klein-Gordon equation near the horizon that satisfies the boundary conditions
imposed by the ``in" modes (\ref{eq:m-ii}), which have been propagated from 
the flat region through regions II
and III. The problem is formally solved using a Bogoliubov transformation
provided $\{u_k^{\rm in}\}$, i.e. the ``in" modes, and the set of modes 
(\ref{eq:phi}), $\{\phi _{k'}(x)\}$,
of the field equation in the interaction region are complete and orthonormal, 
so that we
can write,
 
$$u_{k}^{\rm in}(x)=\sum_{k'}\alpha _{kk'}\left[\phi _{k'}(x)+\beta _{kk'}\phi 
^*_{k'}(x)\right].$$
All the information on the boundary conditions is contained in the Bogoliubov 
coefficients, 
 
\bb \alpha _{kk'}=\langle u_k^{\rm in}(x),\,\phi _{k'}(x)\rangle 
_{\Sigma},\;\;\;\;
\beta _{kk'}=-\langle u_k^{\rm in}(x),\,\phi ^*_{k'}(x)\rangle 
_{\Sigma},\label{eq:alphabetasigma}\ee
where the inner product is evaluated on the boundaries of regions II and III 
with region I and it
is given, see paper I, by,
\bb \langle {\phi}_1
,{\phi}_2\rangle =-i\int dx\, dy\left[\int _{0}^{{\pi}/2}\cos 
^2u\,\left.({\phi}_1
{\buildrel\leftrightarrow\over {\partial}}_{u}{\phi}_{2}^{*})\right|_{v=0}du 
+\int
_{0}^{{\pi}/2}\cos ^2v\,\left.({\phi}_1 {\buildrel\leftrightarrow\over
{\partial}}_{v}{\phi}_{2}^{*})\right|_{u=0}dv\right].\label{eq:escalarsigma}\ee

The coefficients $\alpha ^{{\rm (A)}}_{kk'}$ and $\beta ^{{\rm (A)}}_{kk'}$ 
can be calculated using
(\ref{eq:alphabetasigma}), (\ref{eq:escalarsigma}) and the expressions
(\ref{eq:phi}), (\ref{eq:m-ii}) for the functions $\phi _{k'}(x)$ and 
$u_k^{\rm in}$, respectively. The
superscript
(A) indicates that the values of 
$\phi _{k'}(x)$ are known up to a certain adiabatic order $A$. We have,

$$\alpha ^{{\rm (A)}}_{kk'}=
iM\sqrt{\pi}\, (-1)^m\,\delta (k_x-k_x')\,\delta _{mm'}\,{\cal A} ^{{\rm 
(A)}}_{kk'},\;\;\;\; \beta ^{{\rm
(A)}}_{kk'}= -iM\sqrt{\pi}\,\delta (k_x+k_x')\,\delta _{m(-m')}\,{\cal B} 
^{{\rm (A)}}_{kk'},$$
where ${\cal A} ^{{\rm (A)}}_{kk'}$ and ${\cal B} ^{{\rm (A)}}_{kk'}$ are 
given by, 
 
\bba {\cal A} ^{{\rm (A)}}_{kk'}&=&   
{2\, C\over\sqrt{k_-}}\int _0^{\pi/2}du\,\cos ^2u 
\left[{{\rm e}^{-iA(u)/(2L_2k_-)}\over\cos u }\,{\buildrel\leftrightarrow\over 
{\partial}}_{u} 
\,{\varphi (\eta)\over 1+\sin\xi}\,{{\rm e}^{i\int _0^{\xi}W_1(\xi ') d\xi '^*}
\over\sqrt{W_1(\xi)}}\right]_{v=0}+\nonumber\\
& & {2\, C\over\sqrt{k_-}}\int _0^{\pi/2}dv\,\cos ^2v 
\left[{{\rm e}^{-iA(v)/(2L_1k_+)}\over\cos v }\,{\buildrel\leftrightarrow\over 
{\partial}}_{v} 
\,{\varphi (\eta)\over 1+\sin\xi}\,{{\rm e}^{i\int _0^{\xi}W_1(\xi ') d\xi '^*}
\over\sqrt{W_1(\xi)}}\right]_{u=0},
 \label{eq:AA}\eea

\bba {\cal B} ^{{\rm (A)}}_{kk'}&=&    
{2\, C\over\sqrt{k_-}}\int _0^{\pi/2}du\,\cos ^2u 
\left[{{\rm e}^{-iA(u)/(2L_2k_-)}\over\cos u }\,{\buildrel\leftrightarrow\over 
{\partial}}_{u} 
\,{\varphi (\eta)\over 1+\sin\xi}\,{{\rm e}^{-i\int _0^{\xi}W_1(\xi ') d\xi '^*}
\over\sqrt{W_1(\xi)}}\right]_{v=0}+\nonumber\\
& & {2\, C\over\sqrt{k_-}}\int _0^{\pi/2}dv\,\cos ^2v 
\left[{{\rm e}^{-iA(v)/(2L_1k_+)}\over\cos v }\,{\buildrel\leftrightarrow\over 
{\partial}}_{v} 
\,{\varphi (\eta)\over 1+\sin\xi}\,{{\rm e}^{-i\int _0^{\xi}W_1(\xi ') d\xi '^*}
\over\sqrt{W_1(\xi)}}\right]_{u=0},
\label{eq:BA}\eea
where $\varphi (\eta)$ is a shortened notation for
 
\bb \varphi (\eta)= Y_l^m\left(0,\pi /2-\eta\right)= \left({2l+1\over 
4\pi}{(l-m)!\over (l+m)!}\right)^{1/2}
P_l^m(-\sin\eta),\label{eq:philegendre}\ee 
and where $W_1(\xi)$ is known up to a certain adiabatic order $A$.
 
The problem now is to evaluate these integrals. Since all of them are of the 
type,
$\int dx\, f(x){\rm exp}\left[{i\theta (x)}\right]$,
we search for the intervals of the integration variable where the modulus
of the integration function, $f(x)$, is much larger than the variation of its 
phase $\theta (x)$, i.e 
${\dot\theta}(x)\ll {f}(x)$. In particular, we search for the phase stationary 
points, i.e.
the values of the integration variable for which $\dot\theta (x)=0$.
 
Using that $\xi =u+v$, $\eta =v-u$ and defining star variables $x^*$ by $dx 
^*=dx\, M(1+\sin x)^2/\cos x$,
the phases of the integration functions in (\ref{eq:AA}) and (\ref{eq:BA}) can 
be written, respectively, as
 
\bb {A(u)\over 2L_2k_-}\pm \int _0^u W_1(u')du'^*,\;\;\;\;
{A(v)\over 2L_1k_+}\pm \int _0^v W_1(v')dv'^*,\label{eq:fa1ab}\ee
We can study the problem qualitatively using the value of $W_1(\xi)$ up to 
adiabatic order zero, 
 
\bb W^{\rm (0)}_1={1\over M}\left[k_1^2+{\cos ^2u\over (1+\sin 
u)^4}\left({1\over
L^2}+{2\over 1+\sin u}\right)\right]^{1/2},   \label{eq:W10}\ee
where we have used the notation
$l(l+1)=L^{-2}$.
We have already seen in the previous section, using geometrical optics, that 
the relevant modes involved in any
calculation near the horizon ($\xi =\pi /2$) are those which satisfy 
(\ref{eq:kkkcosu}) in region II (or
(\ref{eq:kkkcosv}) in region III) and this implies (in region II) that the 
labels $L^{-1}$ and
$L_2k_-$, which parametrize the Klein-Gordon modes (\ref{eq:phi}) in the 
interaction region, and the ``in" modes
(\ref{eq:m-ii}) are related. Let us assume that they are proportional, then it 
is easy to see that
expressions (\ref{eq:fa1ab}) have stationary points only when the minus sign 
is involved and these stationary
points are such that
$L^{-2}{\cos ^2u}$ is of the order of $16k_1^2+k_2^2$. This means, in 
particular, that
$L^{-2}{\cos ^2u}$ is of the order of  $k_2^2=m^2$; but for these values, the 
associated Legendre function 
$P_l^m(-\sin u)$ has an oscillating behaviour that we must take into account 
in order to evaluate
the phase stationary points. Note that for the range of values 
$16k_1^2+k_2^2\gg L^{-2}\cos ^2u$, the associated Legendre polinomial 
$P_l^m(-\sin u)$, with $k_2=m$, behaves
as $\cos ^mu$ and in that case we have integrals $\int dx\, f(x){\rm 
exp}\left[{i\theta (x)}\right]$
without stationary points and with $f(u)\sim\cos 
^{m-1}u\ll{\dot\theta}(u)\sim\cos ^{-2}u$, which lead to
negligible contributions.
 
In the following we will find the expressions of the Klein-Gordon
solutions $\phi _k$ in the interaction region for the range of values such 
that $|k_1|L(\cos\xi )^{-1}\ll 1$
and $|k_2|L(\cos\eta )^{-1}\ll 1$. We have seen that solutions $\phi _k$ can 
be separated in two 
functions $\psi (\xi)$ and $\varphi (\eta)$, i.e. $\phi _k(\xi ,\eta)=\psi 
(\xi)\varphi (\eta)$, which satisfy
the differential equations (\ref{eq:e1}) and (\ref{eq:e2}).
The first of these equations can be adiabatically solved  up to any order, see 
(\ref{eq:egam}), and the second
one, although exactly solvable in terms of associated Legendre functions, can 
also be WKB solved
up to any adiabatic order in the range of values such that
${|k_2| L (\cos\eta )^{-1}}\ll 1$. For this we define a new coordinate $\eta 
^*$ by
 $d\eta ^*= d\eta\,{M (\cos\eta )^{-1}}$ 
in (\ref{eq:e2}) and obtain,
 
\bb \varphi _{,\eta ^*\eta ^*}+\omega _2^2\,\varphi=0,\label{eq:varphi*}\ee
where 
\bb \omega _2^2(\eta)\equiv{1\over M^2}\,\left[-k_2^2+{\cos ^2\eta\over
L^2}\right]\equiv{1\over M^2}\,\left[-k_2^2+V_2(\eta)\right],  
\label{eq:omega2}\ee
which is WKB  solved for the values $|k_2|L(\cos\eta )^{-1} \ll 1$ as
 
\bb \varphi (\eta)={C_1\over\sqrt{W_2(\eta)}}\,{\rm e}^{-i\int 
_0^{\eta}W_2(\eta ')d\eta '^*}
+{C_2\over\sqrt{W_2(\eta)}}\,{\rm e}^{i\int _0^{\eta}W_2(\eta ')d\eta '^*}  
\label{eq:varphiW}\ee
where $W_2(\eta)$ can be written up to adiabatic order 4 in exactly the same 
way as
(\ref{eq:Wxi}), 
 
\bb W_2(\eta )=\omega _2 +{{\hat A}_2\over\omega _2^3} +{{\hat B}_2\over\omega 
_2^5} +{{\hat A}_4\over\omega _2^5}
 +{{\hat B}_4\over\omega _2^7} +{{\hat C}_4\over\omega _2^9} +{{\hat 
D}_4\over\omega
_2^{11}},\label{eq:varphiW2}\ee
where the expression of ${\hat A}_2$, ${\hat B}_2,\,\cdots$ are formaly the 
same expressions (\ref{eq:An})
for ${A}_2$, ${B}_2,\,\cdots$, but with $V_1(\xi)$ substituted by $V_2(\eta)$ 
(defined in
(\ref{eq:omega2})), and with the overdot
meaning derivative with respect to $\eta ^*$.
The two arbitrary constants $C_1$ and $C_2$ will be fixed imposing that the 
solution
(\ref{eq:varphiW}) is asymptotically equivalent to the associated Legendre 
function
$P_l^m(-\sin\eta)$ with $L^2=[l(l+1)]^{-1}$ and 
$m=k_2$, for the values $|k_2|L(\cos\eta )^{-1} \ll 1$ (see Appendix A).
 
Before going through, we need to introduce some useful definitions to relate 
the adiabatic
expansions of
$W_1(\xi)$ and $W_2(\eta)$. Let us define,
 
$$q^2={(16k_1^2+k_2^2)^{-1}},\;\;\;\;V^2={16k_1^2\, q^2},\;\;\;\;
U^2={k_2^2\, q^2},$$
note that in practice, and without loss of generality,
we can take $q^2<1$, because the case $q^2\gg 1$, which corresponds to 
$k_1\sim k_2\sim 0$, is equivalent to
consider
$L\ll 1$ in the equations (\ref{eq:e1*}) and (\ref{eq:varphi*}). Note also 
that $0\leq U,\; V\leq 1$,
and that by constuction,
$V^2+U^2=1$.
With these definitions we can write,
 
$$M^2\omega _1^2={V^2\over 16\, q^2}+{\cos ^2\xi\over
(1+\sin\xi)^2}\left({1\over L^2}+{2\over 1+\sin\xi}\right),\;\;\;\; M^2\omega 
_2^2={U^2\over q^2}+{\cos
^2\eta\over L^2}.      $$ 
For parameters that satisfy $|k_1|L(\cos\xi )^{-1} \ll 1$ and $|k_2|L(\cos\eta 
)^{-1} \ll 1$,
we introduce new functions $p(\xi)\equiv Lq^{-1}(\cos\xi)^{-1}$ and 
$p_1(\eta)\equiv Lq^{-1}(\cos\eta)^{-1}$,
so that $p\ll 1$, $p_1\ll 1$.
 
We can now expand (\ref{eq:phi}) and
(\ref{eq:m-ii}) in powers of $p$ and $p_1$ and substitute into (\ref{eq:AA}) 
and (\ref{eq:BA}). Then
the dominant integrals connecting  the ``in" modes between regions II and I 
can be written as
(see Appendix A for details). 
 
\bb 2M\,\int du\, {\rm exp}\left\{\int du\left[p^2\left(
{V^2\over 32}\,(1+\sin u)^4+{U^2\over 2}\right)\left({1\over 
L_2k_-}-{L}\right)+O(p^4)\right]
2M\right\}\label{eq:intdominant}\ee
$$ \times\,\left[p^2\left(
{V^2\over 32}\,(1+\sin u)^4+{U^2\over 2}\right)\left({1\over 
L_2k_-}+{L}\right)+O(p^4)\right]
\simeq \delta\left(L_2k_--{L^{-1}}\right).$$
The dominant integral connecting  the ``in" modes between regions
III and I is obtained from this by changing $L_2k_-$ by $L_1k_+$.
This means that only the $\alpha _{kk'}$ connection coefficients 
(\ref{eq:alphabetasigma}) are relevant in the
propagation of the ``in" modes through the boundaries between regions II and 
III with region I. In other words,
the set of modes $\{\phi _k\}$ of the Klein-Gordon equation for region I 
correspond to the same
initial vacuum state for calculations involving points near the horizon $\xi 
=\pi /2$. This important
result, as we have pointed out in the geometrical optics analysis, can be seen 
as a consequence of the blueshift
that the ``in" modes suffer by a single plane wave. Although $k_-$ (or $k_+$) 
is a continuous label and $L$ is
a discrete label, this should not be a problem since we can use the wave 
packet formalism to discretize
$k_-$ \cite{dor94}.

\section{``Point splitting" regularization technique}
 
In this section we briefly review, for the computational purposes of the 
following sections, the
``point-splitting" regularization technique to calculate the expectation value 
of the stress-energy tensor of a
scalar quantum field in some physical state. The stress-energy tensor of the 
field may be derived by functional
derivation of the action for the scalar field with respect to the metric. When 
the field is massless and the
Ricci tensor is zero it is
\cite{bir82}
 
\bb T_{\mu\nu}=(1-2\xi)\,\phi _{;\mu}\,\phi _{;\nu}+(2\xi-{1\over 
2})\,g_{\mu\nu}\,g^{\alpha\beta}\,\phi
_{;\alpha}\,\phi _{;\beta}-2\xi\,\phi _{;\mu\nu}\,\phi +{1\over 2}\,\xi\, 
g_{\mu\nu} \,
\phi{\nabla}^{\alpha}{\nabla}_{\alpha}\phi
 ,\label{eq:TS1}\ee
where $\xi$ is the coupling parameter of the field to the curvature.
 
To quantize, the
field $\phi$ is promoted into a field operator acting over a given Hilbert 
space $H_{\phi}$
\cite{bir82,wal94}, 
$\phi (x)=\sum _k a_k u_k(x)+a^{\dag}_k u^*_k(x)$,
where $a^{\dag}_k$, $a_k$ are the standard creation and annihilation operators 
and $\{u_k(x)\}$ is a complete
and orthornormal set of solutions of the Klein-Gordon equation (\ref{eq:KG'}). 
Mathematically the field
operator $\phi (x)$ is a point distribution, therefore, the quantum version of 
the stress-energy tensor
(\ref{eq:TS1}) is mathematically pathological because it is quadratic in the
field and its derivatives. One possible way to give  sense  to that expression 
is
to note that the formula (\ref{eq:TS1}) can be formally recovered as,
 
\bb \langle T_{\mu\nu}\rangle (x)=\lim _{x\rightarrow x'}\, {\cal 
D}_{\mu\nu}G^{(1)}(x,x'), \label{eq:limDT}\ee
where $G^{(1)}(x,x')$ is a Green function of the field equation defined as the 
vacuum expectation value of the
anticommutator of the field, and called the {\it Hadamard function}, 
 
\bb G^{(1)}(x,x')=\langle\{\phi (x),\phi (x')\}\rangle =\sum _k 
\left\{u_{k}(x)\, u^*_{k}(x')+u_{k}(x')\,
u^*_{k}(x)\right\}. \label{eq:defHadamard}\ee
As a product of
distributions at different points this is mathematically well defined. The 
differential operator ${\cal
D}_{\mu\nu}$ is given in our case by,
 
\bba{\cal D}_{\mu\nu}&=&
\left(1-2\,\xi\right)\,{1\over 4}\,\left(\nabla _{\mu '}\nabla _{\nu}+\nabla
_{\nu'}\nabla_{\mu}\right)
+  (2\,\xi-{1\over 2})\, g_{\mu\nu}\,{1\over 4}\,\left(
\nabla _{\alpha '}\nabla ^{\alpha}+
\nabla _{\alpha }\nabla ^{\alpha '}\right)-\nonumber\\
& &\xi\,{1\over 2}\,\left(\nabla _{\mu }\nabla _{\nu}+\nabla
_{\mu'}\nabla_{\nu '}\right)
+  \xi\, g_{\mu\nu}\,{1\over 8}\,\left(
\nabla _{\alpha  }\nabla ^{\alpha  }+
\nabla _{\alpha '}\nabla ^{\alpha '}\right).\label{eq:Dopdif}\eea
However, the above differential operation and its limit has no immediate 
covariant meaning because
$G^{(1)}(x,x')$ is not an ordinary function but a {\it biscalar} and the 
differential operator ${\cal
D}_{\mu\nu}$ is {\it nonlocal}, thus we need to deal with the nonlocal 
formalism of {\it bitensors}. 
In Appendix B we give a summary of the main properties of bitensors, how to 
parallel transport from
points $x$ to $x'$ and how to average after the coincidence limit, 
$x\rightarrow x'$, is taken.
 
The above procedure still leads to a divergent quantity since we know that 
even in flat spacetime
$G^{(1)}(x,x')$ has a short-distance singularity and that a ``vacuum" 
substraction has to be performed
to $G^{(1)}(x,x')$ in order to obtain a regularized value. To regularize we 
assume that
$G^{(1)}(x,x')$,  has a short-distance singular structure given by
 
\bb S(x,x')={2\over (4\pi)^2}\Delta ^{1/2}(x,x')\left[-{2\over\sigma
(x,x')}+v(x,x')\ln\sigma(x,x')+w(x,x')\right],      \label{eq:Hada}\ee
where $\sigma (x,x')$ is the {\it geodetic biscalar} (see Appendix B),
$\Delta$ is the {\it Van Vleck-Morette} determinant \cite{dew60}, which is 
singularity free in the
coincidence limit, and where $v(x,x')$ and $w(x,x')$ are biscalars with a 
well-defined coincidence limit for
which we assume the following covariant expansions,
 
\bb v(x,x')=\sum _{l=0}^{\infty}v_l(x,x')\sigma ^l(x,x'),\;\;\;\; w(x,x')=\sum 
_{l=0}^{\infty}w_l(x,x')\sigma
^l(x,x').                      \label{eq:Hvw}\ee 
A Green function expressed in this form is usually called an
{\it elementary Hadamard solution}, the name of which comes from the work of 
Hadamard on the singular structure
for elliptic and hyperbolic second order differential equations. Note, 
however, that this Hadamard
singular structure is not a general feature of any Green function of the 
Klein-Gordon equation. In other words,
although for an extensive range of spacetime and vacuum states, the vacuum 
expectation value of the
anticommutator of the field,
$G^{(1)}(x,x')$, has this singular form, this is not a general property. 
However a theorem
states that if
$G^{(1)}(x,x')$ has the singular structure of an elementary Hadamard solution  
in a neighbourhood of a Cauchy
surface of an arbitrary hyperbolic spacetime, then it has this structure 
everywhere \cite{kay91,ful78}.
As a corollary of this theorem, $G^{(1)}(x,x')$ has this singular structure if 
the spacetime is flat to the past
of a spacetime Cauchy surface, as is the case of our colliding plane wave 
spacetime. This and other
considerations lead to a proposal by Wald \cite{wal94} that any physically
reasonable quantum state must be a {\it Hadamard state}, that is to say, a 
state for which $G^{(1)}(x,x')$
takes the short-distance singular structure of an elementary Hadamard solution.
 
The coefficients $v_i(x,x')$ and $w_i(x,x')$ can
be directly obtained by substitution in the differential equation,
$\Box _x S(x,x')=0$.
Recursion relations for $v_i(x,x')$ and $w_i(x,x')$ are then obtained 
\cite{dew60}. These relations 
uniquely determine
all the $v_i(x,x')$ coefficients  but the coefficients
$w_i(x,x')$ can be written in terms of an arbitrary term $w_0(x,x')$. Up to 
order $\sigma$,
$v(x,x')$ is given by
 
\bb v(x,x')=-a_1(x,x')-{1\over 2}\, a_2(x,x')\,\sigma +\cdots ,   
\label{eq:va1a2}\ee
where  $a_1(x,x')$ and $a_2(x,x')$ are the
Schwinger-DeWitt coefficients \cite{bir82}, which in our case ($R_{\mu\nu}=0$) 
reduce to
 
\bb a_1(x,x')=-{1\over 180}\, {R^{\alpha\beta\gamma}}_{\mu}\,  
R_{\alpha\beta\gamma\nu}\,\sigma ^{\mu}
\sigma ^{\nu}+O(\sigma ^3),\;\;\;\;  a_2(x,x')={1\over 180}\,  
R^{\alpha\beta\gamma\delta}\,
R_{\alpha\beta\gamma\delta}+O(\sigma ).                    
\label{eq:a12}\ee
 
Only the coefficients $v_i(x,x')$ are related to the singular structure of 
$G^{(1)}(x,x')$ in
the coincidence limit, and they are uniquely determined by the spacetime 
geometry. This means that
given any two Hadamard elementary solutions in a certain spacetime geometry, 
both have the same singularity
structure in the coincidence limit; therefore given two vacuum Hadamard 
states, $| 0\rangle$ and
$|{\overline 0}\rangle$, $G^{(1)}(x,x')=\langle 0|\{\phi (x),\phi 
(x')\}|0\rangle$ and 
${\overline G}^{(1)}(x,x')=\langle {\overline 0}|\{\phi (x),\phi 
(x')\}|{\overline 0}\rangle$, they have the
same singular structure. Their finite parts, however, may differ because the 
two vacuum states are related to
different boundary conditions, which are global spacetime features. 
Mathematically this comes from the fact that
the term $w_0(x,x')$ in the elementary Hadamard solution is totally arbitrary, 
fixing $w_0(x,x')$ we fix
a particular boundary condition. This suggests a possible renormalization 
procedure
\cite{adl77,wal78,wal94}: we can eliminate
the non-physical divergences of any $G^{(1)}(x,x')$ without alterations in the 
particular physical boundary
conditions by subtracting an elementary Hadamard solution with the particular 
value $w_0(x,x')=0$, which
is the value that
corresponds to the flat space case. In other words, we define the following 
regularized biscalar,
 
\bb G_B^{(1)}(x,x')=G^{(1)}(x,x')-\left.S(x,x')\right|_{w_0=0}.\label{eq:GB}\ee
Then by means of $G_B^{(1)}(x,x')$ we can construct a 
$\left<T_{\mu\nu}^B\right>$ by differentiation with the
nonlocal operator (\ref{eq:Dopdif}). 
 
This regularization procedure, however, fails to give a covariantly conserved 
stress-energy tensor since it can
be seen  \cite{wal78} that for a massless conformal
scalar field (i.e.
$\xi =1/6$),
 
\bb \nabla ^{\nu} \langle T_{\mu\nu}^B\rangle
={1\over 4}\,\lim _{x\rightarrow x'}\nabla _{\mu}\,
\left(\Box _{x'}+{1\over 6}\, R(x')\right)\,
G^B(x,x')={1\over 64\pi ^2}\, \nabla _{\mu}\, a_2(x),
  \label{eq:anomalia}\ee 
where $a_2(x)$ is the coincidence limit of the Schwinger-DeWitt coefficient
$a_2(x,x')$ in (\ref{eq:a12}). In particular
for a spacetime with null curvature $R=0$, such as our colliding spacetime, 
this is true for any coupling. 
Thus to ensure covariant conservation, we must introduce an additional 
prescription:
 
\bb \langle T_{\mu\nu}(x)\rangle =\langle T_{\mu\nu}^B (x)\rangle
- {a_2(x)\over 64\pi ^2}\, g_{\mu\nu}. \label{eq:GBT}\ee
Note that this last term is responsible for the trace anomaly in the conformal 
coupling, because even though
$\langle T_{\mu\nu}^B (x)\rangle$ has null trace when $\xi =1/6$, the trace of 
$\langle T_{\mu\nu}(x)\rangle$
is given by
$\langle T^{\mu}_{\mu}\rangle =
- {a_2(x)(16\pi ^2)^{-1}}$.
 
The regularization prescription just given satisfies the well known four 
Wald's axioms
\cite{wal76,wal77-78b,chr75,wal94}, a set of properties that any physically 
reasonable expectation value of the
stress-energy tensor of a quantum field should satisfy. There is still an 
ambiguity in this prescription since
two independent conserved local curvature terms, which are quadratic in the 
curvature, can be added to this
stress-energy tensor. This two-parameter ambiguity, however, cannot be 
resolved within the limits of the
semiclassical theory, it may be resolved in a complete quantum theory of gravity
\cite{wal94}. Note, however, that in some sense this ambiguity does not affect 
the knowledge of the matter
distribution because a tensor of this kind belongs properly to the left hand 
side of Einstein equations, i.e.
to the geometry rather than to the matter distribution.

\section{Vacuum expectation value of the stress-energy tensor for a single
plane wave}
 
A simple nontrivial example that nicely ilustrates the  point splitting 
technique for
the regularization of the vacuum expectation value of the stress-energy tensor 
is the particular
case of a single gravitational plane wave, i.e. the case of regions II and III,
$\langle T_{\mu\nu}\rangle _{II}$ and $\langle T_{\mu\nu}\rangle _{III}$. The 
result is not new,
it is known that the expectation value of the stress-energy tensor
in the ``in" state, i.e. the vacuum state at the flat region before the 
passing of the wave, is
exactly zero \cite{gib75,vil77}. Let us start with the construction of the
Hadamard function $G^{(1)}(x,x')$, (\ref{eq:defHadamard}), if the vacuum state 
is the ``in" vacuum state, we
have
 
\bban G^{(1)}(x,x')&=&\sum _k u_k(x)\, u_k^*(x')+{\rm c.c.}={\sec u\,\sec 
u'\over 2 (2\pi)^3}
\,\int _0^{\infty} {dk_-\over k_-}\, {\rm e}^{-i2L_2k_- \, 
(v-v')}\,\times\nonumber\\
& &\hskip -1.5 truecm\int _{-\infty}^{\infty} dk_x\, 
{\rm e}^{ik_x\, (x-x')-i2L_1\,\left[f(u)-f(u')\right]\, k_x^2/k_-}\,
\int _{-\infty}^{\infty} dk_y\, 
{\rm e}^{ik_y\, (y-y')-i2L_1\,\left[g(u)-g(u')\right]\, k_y^2/k_-}\; +{\rm 
c.c.},\eean
where $u_k$ are the ``in" modes (\ref{eq:m-ii}) of region II. It is easy to 
compute
analytically the above ``mode sum" since only Gaussian integrals are involved, 
the result is
 
\bba G^{(1)}(x,x')&=&
{2\over 4 M^2 (2\pi)^2}\,{\sec u\,\sec u'\over 
\sqrt{\left[f(u)-f(u')\right]\left[g(u)-g(u')\right]}}
\,\times\nonumber\\
& & \left\{4 M^2(v-v')+\left({(x-x')^2\over f(u)-f(u')}+{(y-y')^2\over 
g(u)-g(u')}\right)\right\}^{-1}.
\label{eq:GHaPW}\eea

We can now proceed with the point splitting formalism by setting the points 
$x$ and $x'$ at the
endpoints of a non null geodesic parametrized by its proper distance. Let us 
denote by
$\bar\tau$ the midpoint on the geodesic, which lies at equal proper distance 
from $x$ and $x'$, i.e,
 
$$x^{\mu}=x^{\mu}({\bar\tau} +\epsilon),\;\;\;\; {\bar 
x}^{\mu}=x^{\mu}({\bar\tau}),\;\;\;\;
x'^{\mu}=x^{\mu}({\bar\tau} -\epsilon),$$ 
where one should be aware that the notation $x$, $x'$ may be a little
ambiguous because it makes reference to the points
$(u,v,x,y)$,
$(u',v',x',y')$ and also to their third components.
 
Now we can easily solve the geodesic equations, since the momenta $p_x$, $p_y$ 
and $p_v$ are three constants of
the motion. The geodesics are a family
of curves parametrized by $p_x$, $p_y$ and $p_v$,
 
\bb x=x(u)=x(0)-2\, M^2\,{p_x\over p_v}\, f(u),\;\;\;\; y=y(u)=y(0)-2\, 
M^2\,{p_y\over p_v}\, g(u),
\label{eq:geoxy}\ee
 
\bb v=v(u)=v(0)+ M^2\,{p^2_x\over p^2_v}\, f(u)+M^2\,{p^2_y\over p^2_v}\, g(u)+
{M^2\over p^2_v}\,\int _0^u du'\, (1+\sin u)^2, \label{eq:geov}\ee
where $f(u)$ and $g(u)$, are given in (\ref{eq:fg}). If we substitute these 
curves into (\ref{eq:GHaPW}) using
that
$x'=x(u')$, $y'=y(u')$, $v'=v(u')$ and take $u$, $\bar u$, $u'$ as
$u=u({\bar\tau} +\epsilon)$, ${\bar u}=u({\bar\tau})$, $u'=u({\bar\tau} 
-\epsilon)$,
then we only need to expand (\ref{eq:GHaPW}) in powers of $\epsilon$ up to 
order $\epsilon ^2$. The
calculation is simple but tedious and finally one obtains,  
 
\bb G^{(1)}(x,x')=-{1\over 4 \pi ^2} {1\over\sigma} - {p_v ^4\over 160\, 
M^8\,\pi ^2}\,
{\sigma\over (1+\sin {\bar u})^{10}}. \label{eq:G-PW1}\ee
where $\sigma =\Sigma s^2/2=2\,\epsilon ^2\Sigma$ and $s$ is the proper 
distance along the geodesic between
$x$ and $x'$, see Appendix B.
 
The regularization follows from the subtraction of the elementary Hadamard 
solution (\ref{eq:Hada})
$$\left. S(x,x')\right|_{w_0}=-{1\over 4 \pi ^2} {1\over\sigma},$$
since the biscalars $a_2(x,x')$ and $a_1(x,x')$ are zero up to order $\sigma$. 
Note that, in particular, 
there is no trace anomaly because $a_2(x)=0$, thus after regularization there 
remains a term proportional to
 
$$p_v^4\,\sigma = 4{\sigma _{\bar v}^4\over \sigma}=4{\sigma 
_{v}^4\over\sigma}+O(\epsilon ),$$
where we have used that $\sigma ^{\mu}=2\epsilon p^{\mu}$ being $p^{\mu}$ a 
tangent vector to the geodesic
connecting $x$ with $x'$ such that $p^{\mu}p_{\mu}=\Sigma$. Under application 
of the
differential operator (\ref{eq:Dopdif}) (details are given in section 9) one 
easily gets an expression which
involves factor terms of the kind $p_vp_v$, $p_vp_vp_vp_{\mu}$ and 
$p_vp_vp_vp_vp_{\mu}p_{\nu}$. These terms
are path dependent in the limit of coincidence $x\rightarrow x'$ therefore we 
must
introduce the elementary averaging procedure of Appendix B, which gives,
 
$$\langle p_vp_v \rangle \propto g_{vv}=0,\;\;\;\;
\langle p_vp_vp_vp_{\mu} \rangle \propto g_{vv}=0,\;\;\;\; 
\langle p_vp_vp_vp_vp_{\mu}p_{\nu} \rangle \propto g_{vv}=0.$$
Thus we recover from (\ref{eq:limDT}) the known result that
the vacuum expectation value of the stress-energy tensor of a quantum scalar 
field, in a single plane wave
region is exactly zero, $\langle T_{\mu\nu}\rangle _{II}=0$, i.e. exact 
gravitational plane waves do not
polarize the vacuum.

\section{Hadamard function in the interaction region}
 
Here we calculate the Hadamard function $G^{(1)}(x,x')$ in the 
``in" vacuum state in the interaction region, near the horizon. As we have 
seen in section 5, due to the
blueshift of the ``in" modes towards the horizon, the $u_k^{\rm in}$ with 
large energy values (which are the
relevant modes in calculations near the horizon)
determine the same vacuum state as the
modes $\phi _k$, (\ref{eq:phi}). This means that near the horizon the Hadamard 
function in this vaccum state
can be written as,
 
\bb G^{(1)}(x,x')=\sum _k\left\{u_k^{\rm in}(x)\,u_k^{{\rm in}*}(x')\right\}\; 
+{c.c.}=\sum _k\left\{\phi
_k(x)\,\phi ^*_k(x')\right\}\; +{c.c.}    
\label{eq:G(1)}\ee
Since we do not have an exact expression for the solutions $\phi _k$ we need to
work up to a certain adiabatic approximation. This means that we have the 
inherent ambiguity of where to
cut the adiabatic series. Fortunately the boundary conditions of the problem 
impose that the
asymptotic value for the ``in" modes on the horizon is given by,
${\rm exp}\left({-i|k_x|\xi ^*}\right)$, see paper I for details. This is just 
the form for the ``in" modes
that we used in paper I to calculate the production of particles, and is 
equivalent to cut the
adiabatic series at order zero. Although for the particle production problem 
it was enough to cut at order
zero, it is not sufficient for the calculation of the vacuum expectation value 
of the stress-energy tensor.
This is because
$G^{(1)}$ at order zero does not reproduce the short-distance singular 
structure of a Hadamard elementary
solution (\ref{eq:Hada}) in the coincidence limit $x\rightarrow x'$. The 
smallest adiabatic
order which we need, to recover the singular structure of $G^{(1)}$, is order 
four.
 
In the mode sum of (\ref{eq:G(1)}) we use the shortened notation $\sum 
_k\equiv \int
^{\infty}_{-\infty}dk_x\,\sum _{l=0}^{\infty}\,\sum _{m=-l}^{m=l}, $  and we 
note from 
(\ref{eq:phi}) that the sum over $m$ is
straightforward using the property of the spherical harmonics,
 
$$\sum _{m=-l}^{m=l} Y_l^m(\pi /2-\eta,\, y/M) Y_l^{m*}(\pi /2-\eta ',\, 
y'/M)={2l+1\over 4\pi}\,
P_l(\cos\Theta), $$
where $P_l(x)$ is a  Legendre polynomial of order $l$ and $\Theta$ is the 
angle  between the points
$(\theta,\;\varphi)$ and
$(\theta ',\;\varphi ')$ on the sphere, i.e,
$\cos\Theta =\cos\theta\,\cos\theta '+\sin\theta\,\sin\theta 
'\,\cos\left(\varphi -\varphi
'\right)$. 
The limit $(\theta '\;\varphi ')\rightarrow (\theta\;\varphi)$, or 
equivalently $\Theta\rightarrow 0$,
can be taken right away, it is not singular and 
it is just related to the spherical symmetry of the horizon, provided the 
coordinate $y$ is cyclic.
Then, since
$P_l(1)=1$,
            
\bba  G^{(1)}(x,x')&=&{C^2\over \pi}\,{1\over (1+\sin\xi)(1+\sin\xi ')}\,\int 
^{\infty}_{-\infty}dk_x\,
{\rm e}^{ik_x (x-x')}\,\times \nonumber\\
& &\sum _{l=0}^{\infty}\,(2l+1)\,{1\over\sqrt{W_1(\xi)W_1(\xi ')}}\,
{\rm e}^{-i\int _{\xi '}^{\xi}W_1(\xi '')d\xi ''^*}.                         
\label{eq:G(1)2}\eea
 
Let us start now with the point splitting procedure. We assume that the points 
$x$ and $x'$ are connected by a
non null geodesic in such a way that they are at the same proper distance 
$\epsilon$ from a third midpoint
$\bar x$. We parametrize the geodesic by its proper distance $\tau$ and denote 
the end points by
$x$ and $x'$ (it should not be confused with the third
component of $(\xi ,\;\eta ,\; x,\; y)$). Then we expand,
 
\bb x^{\mu}=x^{\mu}(\tau)=x^{\mu}({\bar\tau}+\epsilon)=x^{\mu}_0+\epsilon\, 
x^{\mu}_1+ 
{\epsilon ^2\over 2!}\, x^{\mu}_2+\cdots , \label{eq:xm}\ee 
\bb x'^{\mu }=x^{\mu }(\tau ')=x^{\mu '}({\bar\tau}-\epsilon)=x^{\mu 
}_0-\epsilon\, x^{\mu }_1+ 
{\epsilon ^2\over 2!}\, x^{\mu }_2-\cdots , \label{eq:xm'}\ee 
where we have defined,
 
\bb x^{\mu}_0\equiv x^{\mu}(\tau),\;\;\; x^{\mu}_1\equiv \left.{d
x^{\mu}\over d\tau}\right|_{\bar\tau}, \;\;\; x^{\mu}_2\equiv\left.{d^2
x^{\mu}\over d\tau ^2}\right|_{\bar\tau}, \cdots ,                  
\label{eq:xn}\ee
i.e. the subindex of a given coordinate indicates the number of derivatives 
with respect to $\tau$ at the point
$\bar\tau$.
 
Using these expansions in $\epsilon$ we can write $\int _{\xi '}^{\xi}W_1(\xi 
'')d\xi ''^*$,
$\left[W_1(\xi )W_1(\xi ')\right]^{-1/2}$ and the prefactor 
$\left[(1+\sin\xi)(1+\sin\xi ')\right]^{-1}$
in (\ref{eq:G(1)2}) up to
$O(\epsilon ^4)$. Note also that $x-x'=2\,\epsilon\, x_1 +({\epsilon ^3/ 3})\, 
x_3+({\epsilon ^5/ 60})\,
x_5+O(\epsilon ^7)$. With these expressions we can expand the exponential term 
in
(\ref{eq:G(1)2}) in powers of $\epsilon$ as
 
$$ {\rm e}^{-i\int _{\xi '}^{\xi}W_1(\xi '')d\xi ''^*+ik_x\, (x-x')}=
{\rm e}^{-i\epsilon\left(\delta _1\,\omega-\delta _2\,k_x\right)}\,\times\, 
\left(
1+\cdots\right),$$
where we preserve from the expansion the
zero adiabatic terms because they will be useful as new integration variables. 
These terms have been included
in the coefficients
$\delta _1$ and $\delta _2$ as,
 
\bb  \delta _1= 2\xi _1^*+{\epsilon ^2\over 3}\xi _3^*+{\epsilon ^4\over 
60}\xi _3^* ,\;\;\;\; \delta _2=
2x_1+{\epsilon ^2\over 3}x_3 +{\epsilon ^4\over 60}x _3^* .        
\label{delta12}\ee 
 
Since the number of derivatives $d/d\xi ^*$ determine the adiabatic order, we 
introduce a new
parameter $T$, which will indicate the adiabatic order, then at the end
of calculations we will take $T=1$. We will also denote,
 
$$\omega\equiv \omega _1({\bar\xi}),\;\;\;\; V_l\equiv V_1({\bar\xi}),\;\;\;\; 
W\equiv W_1({\bar\xi}),\;\;\;\; 
{\dot W}\equiv \left.{dW_1(\xi)\over d\xi ^*}\right|_{\bar\xi}.$$
Recall that from section 3 we know the function $W_1$ up to adiabatic order 
four, thus from (\ref{eq:Wxi}) we
have now,
 
$$W=\omega +T^2\,\left({A_2\over\omega ^3}+{B_2\over\omega ^5}\right)+
T^4\,\left({A_4\over\omega ^5}+{B_4\over\omega ^7}+{C_4\over\omega 
^9}+{D_4\over\omega ^{11}}\right),    $$
where the coefficients $\omega _1$, $A_n$, $B_n$, $C_n$, are given in 
(\ref{eq:omegaxi}) and (\ref{eq:An}).
 
We substitute these expansions in (\ref{eq:G(1)2}) and separate the different 
adiabatic terms by expanding
in powers of the parameter $T$. The process is quite simple but tedious and 
the result is,
 
\bba  G^{(1)}(x,x')&=&{C^2\over \pi}\,
\left[{1\over (1+\sin{\bar\xi})^2}+
\epsilon ^2\,\left(\xi _1^{*2}\,{(3-\sin{\bar\xi})\,\cos ^2{\bar\xi}\over 
M^2(1+\sin{\bar\xi})^{7}}-
\xi _2^*\,{\cos ^2{\bar\xi}\over M(1+\sin{\bar\xi})^{5}}\right)
+O(\epsilon ^4)\right] \nonumber\\
& &\hskip -2.5 truecm\times\,\int ^{\infty}_{-\infty}dk_x\,\sum 
_{l=0}^{\infty}\,(2l+1)\, {\rm
e}^{-i\epsilon\left(\delta _1\,\omega-\delta _2\,k_x\right)}\, 
\left[{\cal A}_0+i\epsilon\, {\cal A}_1+\epsilon ^2\, {\cal A}_2+
i\epsilon ^3\, {\cal A}_3+\epsilon ^4\, {\cal A}_4 +O(\epsilon 
^5)\right].\label{eq:G(1)3}\eea
where the coefficients ${\cal A}_n$ are given in Appendix C.
 
Let us now proceed to the integration in (\ref{eq:G(1)3}), evaluating first 
the summation over the discrete
parameter $l$. It is convenient to transform the above summation into an 
integral by means of the
Euler-Maclaurin summation formula \cite{abr72}, 
 
\bb   \sum _{l=0}^{n-1}F(l)=\int _0^n F(x)\, dx+{1\over 
2}\,\left[F(0)-F(n)\right]
-{1\over 12}\,\left[F'(0)-F'(n)\right]
+{1\over 720}\,\left[F'''(0)-F'''(n)\right]+\cdots ,                   
\label{eq:EM0}\ee
where ${F}'=dF(l)/dl$. In our case $n\rightarrow\infty$ and
$F^{(i)}(l)$ is given from (\ref{eq:G(1)3}) as,
$F^{(i)}(l)\propto{\rm exp}\left({-i\epsilon\delta _1\omega}\right)$.
Note that in the limit $n\rightarrow\infty$
with finite geodesic distance $\epsilon$, the phase in $F^{(i)}(n)$ is highly 
oscillating but the
modulus is either aproaching to zero or is finite. In other words, the 
contribution at $n\rightarrow\infty$ is
negligible, this can be mathematically reproduced by introducing a small 
negative imaginary
part in the geodesic splitting as,
$\epsilon\rightarrow\epsilon -i0^+$,
which allows us to regularize  from the very beginning:
$\lim _{n\rightarrow\infty}F^{(i)}(n)=0$.
Therefore we write,
 
\bb  \sum _{l=0}^{n-1}F(l)=\int _0^n F(x)\, dx+{1\over 2}\, F(0)
-{1\over 12}\, F'(0)
+{1\over 720}\, F'''(0)+\cdots \equiv\int _0^n F(x)\, dx+\left.{\cal 
L}F(x)\right|_{x=0}
,                   
\label{eq:EM}\ee
where the  definition of the differential operator $\cal L$ should be clear 
from the last
identity. Then  we can substitute the summation in (\ref{eq:G(1)3}) for an 
integral over a formally continuous
parameter $l$. This integral can be further simplified by using a new 
integration variable $u$,
 
\bb  {u\, M^{-1}}=\omega=M^{-1}\left[k_1^2+V_l\right]^{1/2},\;\;\;\;
V_l=a({\bar\xi})\, l(l+1)+c({\bar\xi}), 
\label{eq:canvi-u}\ee  
where we define, see (\ref{eq:omegaxi}),
$a({\bar\xi})=\cos ^2{\bar\xi} (1+\sin{\bar\xi})^{-4}$, $c({\bar\xi})=2\cos 
^2{\bar\xi} (1+\sin{\bar\xi})^{-5}$.
Then 
$2u\, du=a({\bar\xi})\, (2l+1)\, dl$,
and,
 
$$l(l+1)={u^2-\Omega ^2\over a({\bar\xi})},\,\;\;\;\;\Omega\equiv 
k_1^2+c({\bar\xi}),$$
where for simplicity we introduce a new dimensionless variable $\Omega$. With 
this one
can directly substitute the derivatives  $d^iV_l/d{\bar\xi} ^{*\, i}\equiv 
V^{(i)}_l$ which appear in the
coefficients
${\cal A}_n$ by,
 
\bb V^{(i)}_l={1\over M^2}\left[{a^{(i)}({\bar\xi})\over a({\bar\xi})}\, 
(u^2-\Omega
^2)+c^{(i)}({\bar\xi})\right].\label{eq:diV}\ee 
With all this (\ref{eq:G(1)3}) takes the form,
 
$$  G^{(1)}(x,x')={C^2\over \pi}\,\left[{1\over (1+\sin{\bar\xi})^2}+
\epsilon ^2\,\left(\xi _1^{*2}\,{(3-\sin{\bar\xi})\,\cos ^2{\bar\xi}\over 
M^2(1+\sin{\bar\xi})^{7}}-
\xi _2^*\,{\cos ^2{\bar\xi}\over M(1+\sin{\bar\xi})^{5}}\right)
+O(\epsilon ^4)\right]\,\times $$ 
$$\int ^{\infty}_{-\infty}dk_x\,{\rm e}^{i\epsilon\delta _2\,k_x}
\int_{\Omega}^{\infty}\, {2\, u\over a({\bar\xi})}\, du\,  {\rm
e}^{-i\epsilon\delta _1\, u/M}
\left[{\cal A}_0+i\epsilon\, {\cal A}_1+\epsilon ^2\, {\cal A}_2+
i\epsilon ^3\, {\cal A}_3+\epsilon ^4\, {\cal A}_4 +O(\epsilon ^5)\right]+$$
\bb +{\rm e}^{i\epsilon\delta _2\,k_x}\, {\cal L}\,\left[(2l+1)\, 
\left({\cal A}_0+i\epsilon\, {\cal A}_1+\epsilon ^2\, {\cal A}_2+
i\epsilon ^3\, {\cal A}_3+\epsilon ^4\, {\cal A}_4 +O(\epsilon ^5)\right)\,
{\rm e}^{-i\epsilon\delta _1\,\omega}\right]_{l=0},
\label{eq:G(1)5}\ee 
where the functions ${\cal A}_n$ are now written in powers of $u^{-1}$ and 
$\Omega$. 
The integrals that we have in 
(\ref{eq:G(1)5}) are of the type,, 
 
\bb {\cal I}^m_n=\int _{-\infty}^{\infty}dk_1\,\int _{\Omega}^{\infty} 
{\rm e}^{-i\epsilon\left\{\delta _1\, u/M -\delta _2\, k_x\right\}}\, 
{\Omega ^m\over u^n}\, du  ,\label{eq:calimn}\ee
which can be easily integrated over $u$ using the following identities,
 
$$\int _{\Omega}^{\infty} {\rm e}^{-i\epsilon\delta _1 u/M}\, u\,du=
-\left({M^2\over\epsilon ^2\,\delta _1^2}+i{M\,\Omega\over\epsilon \,\delta 
_1}\right)\,
{\rm e}^{-i\epsilon\,\delta _1\,\Omega /M}, \;\;\;\; \int _{\Omega}^{\infty} 
{\rm e}^{-i\epsilon\delta _1 u/M}\,
du=-{iM\over \epsilon\delta _1}\, {\rm e}^{-i\epsilon\,\delta _1\,\Omega /M},$$
 
\bban\int _{\Omega}^{\infty} {\rm e}^{-i\epsilon\delta _1 u/M}\, 
{du\over u^n}&=&
-\left(-i{\epsilon\,\delta _1\over M}\right)^{n-1}\,{\rm 
E}_i(-i{\epsilon\,\delta _1\,\Omega /M})+\\
& &
{\rm e}^{-i\epsilon\,\delta _1\,\Omega /M}\,{1\over (n-1)!}\,{1\over\Omega 
^{n-1}}\,\sum
_{k=0}^{n-2}\, 
\left(-i{\epsilon\,\delta _1\,\Omega\over M}\right)^{k}\, (n-2-k)!\; ;\;\; n>0.
\eean
where we have used a small negative imaginary part in $\epsilon$, i.e. 
$\epsilon -i0^+$, and where ${\rm
E}_i(ix)$ is the integral-exponential function with imaginary argument 
\cite{gra80}. The remaining
integrals are,
 
$$ {\cal I}^{-m}_0\equiv {\cal I}_{-m}=\int _{-\infty}^{\infty} 
{\rm e}^{-i\epsilon\left(\delta _1\,\Omega -\delta _2\, k_1\right)/M}\, 
{dk_1\over\Omega ^m}=O(\epsilon ^{(m-1)})\, \ell, 
\;\;\;\; m> 0,$$
 
$$ {\cal I}^{m}_0\equiv {\cal I}_m=\int _{-\infty}^{\infty} 
{\rm e}^{-i\epsilon\left(\delta _1\,\Omega -\delta _2\, k_1\right)/M}\,\Omega 
^m\, dk_1=
O(\epsilon ^{-(m+1)})+
\left\{\begin{array}{ll}
O(\epsilon )\, \ell                        & m\geq 0\;\;{\rm even},\\
\left[1+O(\epsilon ^2)\right]\, \ell       & m\geq 0\;\;{\rm odd},
\end{array}\right.
$$
 
$$ {\cal IE}_m=\int _{-\infty}^{\infty} 
{\rm e}^{i\epsilon\,\delta _2\, k_1/M}\,
{\rm E}_i(-i{\epsilon\,\delta _1\,\Omega /M})\,\Omega ^{m}\, dk_1=
O(\epsilon ^{-(m+1)})+O(\epsilon )\, \ell,
\;\;\;\; m\geq 0\;\;{\rm even},$$
where $\ell$ is a logarithmic term defined
as $\ell=\gamma+\ln (\epsilon /M)$, where $\gamma$ is Euler's constant. These 
integrals can be calculated using
the indications of  Appendix D. 
 
Many of the terms in (\ref{eq:G(1)5}), however, are not relevant for our
calculation because they give results of order beyond
$\epsilon ^2$. In fact, note that the series
${\cal A}_0+i\epsilon\, {\cal A}_1+\epsilon ^2\, {\cal A}_2+
i\epsilon ^3\, {\cal A}_3+\epsilon ^4\, {\cal A}_4 +O(\epsilon ^5)$
has dimension of $M^{-1}$, $\Omega$ and $u$ are
dimensionless variables, $T$ has dimensions of $M^{-1}$, and the geodesic 
coefficients $d^n x^{\mu}/d\tau ^n$
have dimensions of $M^{-(n-1)}$. Now a simple dimensional analysis shows that 
the singularity $\sigma
^{-2}$ in the Hadamard function is recovered with the single term
${\cal A}_0$ up to adiabatic order zero only, the logarithmic singularity, 
i.e. $\ln\sigma$, is
recovered also with the single term ${\cal A}_0$,
but now  up to adiabatic order two, and the singularity $\sigma\ln\sigma$ is
recovered with the  terms ${\cal A}_0$,
${\cal A}_1$, ${\cal A}_2$ and ${\cal A}_3$ up to adiabatic order four.

We can now substitute in (\ref{eq:G(1)5})  the values of the integrals given 
in Appendix D,
the values for the functions $a$ and $c$ given in (\ref{eq:canvi-u}), the 
expression 
(\ref{delta12}), for the coefficients $\delta _1$, $\delta _2$, 
and the relation between the geodesic coefficients $\xi _n^*$ i $x_n$ in terms 
of
$\xi ^*_1$ and $x_1$ (see Appendix E). Also we use the identities, see 
Appendix B, $\sigma = 2\epsilon
^2\Sigma$, ${\sigma}^{\bar\mu}=2\epsilon p^{\bar\mu}$, 
$p^{\bar\mu}p_{\bar\mu}=\Sigma$, where 
${\sigma}^{\bar\mu}$ is the 
geodesic tangent vector in the midpoint $\bar x$ with modulus the
proper distance on the geodesic between $x$ and $x'$. In particular,
${\sigma}^{{\bar x}}=2\epsilon\, x_1$.
Finally, after a rather tedious calculation we obtain the expression for the 
Hadamard function, 
 
$$ G^{(1)}(x,x')=-{4\, M^2\, C^2\over\pi\sigma}+{\bar A}+\sigma\,{\bar B}+
{C}_{{\bar x}{\bar x}}\,{\sigma}^{\bar x}{\sigma}^{\bar x}+
D_{{\bar x}{\bar x}{\bar x}{\bar x}}\,
{{\sigma}^{\bar x}{\sigma}^{\bar x}{\sigma}^{\bar x}{\sigma}^{\bar 
x}\over\sigma},$$
where the normalization constant $C$, by comparison with (\ref{eq:Hada}), is 
necessarily 
$C^2=(16M^2\pi)^{-1}$ and the explicit expressions for ${\bar A}$, ${\bar B}$, 
${C}_{{\bar x}{\bar x}}$ and
$D_{{\bar x}{\bar x}{\bar x}{\bar x}}$ are given up to adiabatic order four in 
Appendix C.
Note that, as it is expected on general grounds, there is no logarithmic 
singularity. In fact the
coefficient for the logarithmic term is,
 
$$-{2\over 3}+2\,{c\over a}+{M^2\over 3}\,{{\dot a}^2\over a^3}
-{M^2\over 3}\,{{\ddot a}\over a^2},$$
which is identicaly zero as one can easily see by direct substitution of the 
value for the 
functions $a$ and $c$, in (\ref{eq:canvi-u}).
 
According to (\ref{eq:GB}), the Hadamard function can be regularized using the 
elementary
Hadamard solution (\ref{eq:Hada}), which in our case  reduces simply to,
 
$$\left. S(x,x')\right|_{w_0}=-{1\over 4\pi ^2\sigma},$$
since the coefficient of the logarithmic divergence, see (\ref{eq:va1a2}),
exactly cancells up to order $\sigma$ for a Schwarzshild geometry (the 
interaction region is
locally isometric to Schwarzshild). In fact, for our geometry and using 
(\ref{eq:a12})
 
$$-a_1-{1\over 2}\, a_2\,\sigma =
{1\over 180}\, {R^{\alpha\beta\gamma}}_{\mu}\, 
R_{\alpha\beta\gamma\beta}\,\sigma ^{\mu}\,\sigma ^{\nu}-
{1\over 4}\,{1\over 180}\, R^{\alpha\beta\gamma\delta}\, 
R_{\alpha\beta\gamma\delta}\,\sigma =0,$$
Finally the regularized expression for
$G^{(1)}(x,x')$ correctly normalized up to order $\epsilon ^2$ is,
 
\bb G^{(1)}_B(x,x')={\bar A}+\sigma\,{\bar B}+
{C}_{{\bar x}{\bar x}}\,{\sigma}^{\bar x}{\sigma}^{\bar x}+
D_{{\bar x}{\bar x}{\bar x}{\bar x}}\,
{{\sigma}^{\bar x}{\sigma}^{\bar x}{\sigma}^{\bar x}{\sigma}^{\bar 
x}\over\sigma}.\label{eq:G(1)fR}\ee
From this expression we can directly read the regularized mean square field in 
the ``in" vacuum state as
$\langle\phi ^2\rangle ={\bar A}$.

\section{Expectation value of the stress-energy tensor}

To calculate the vacuum expectation value of the stress-energy
tensor near the horizon we have to apply the differential operator 
(\ref{eq:Dopdif}) to
(\ref{eq:G(1)fR}). As we have already pointed out, this is not straightforward 
because we
work with nonlocal quantities. Note first that the operator (\ref{eq:Dopdif}) 
acts over bitensors which
depend on the end points
$x$ and $x'$, but the expression (\ref{eq:G(1)fR}) for $G^{(1)}_B$ depends on 
the midpoint $\bar x$. It means,
therefore, that we need to covariantly expand (\ref{eq:G(1)fR}) in terms of 
the endpoints $x$ and $x'$.
Consider the general expression, 
 
\bb G^{(1)}_B(x,x')={\bar A}+\sigma\,{\bar B}+
{C}_{{\bar\alpha}{\bar\beta}}\,{\sigma}^{\bar\alpha}{\sigma}^{\bar\beta}+
D_{{\bar\alpha}{\bar\beta}{\bar\gamma}{\bar\delta}}\,
{{\sigma}^{\bar\alpha}{\sigma}^{\bar\beta}{\sigma}^{\bar\gamma}{\sigma}^{\bar%
\delta}\over\sigma}
,\label{eq:GABCD}\ee
where it is understood that ${\bar A}$, ${\bar B}$, 
${C}_{{\bar\alpha}{\bar\beta}}$, 
$D_{{\bar\alpha}{\bar\beta}{\bar\gamma}{\bar\delta}}$ are functions
that depend on the endpoints $x$, $x'$ but are evaluated at the midpoint $\bar 
x$. Then, following the
formalism of Appendix B, we can expand in the neighbourhood of $x$ as,
 
\bb G^{(1)}_B(x,x')={\bar A}+\sigma\,{\bar B}+
{{\bar g}^{\bar\alpha}}_{\alpha}\,{{\bar g}^{\bar\beta}}_{\beta}\,
{C}_{{\bar\alpha}{\bar\beta}}\,{\sigma}^{\alpha}{\sigma}^{\beta}+
{{\bar g}^{\bar\alpha}}_{\alpha}\,{{\bar g}^{\bar\beta}}_{\beta}\,{{\bar
g}^{\bar\gamma}}_{\gamma}\,{{\bar g}^{\bar\delta}}_{\delta}\,
D_{{\bar\alpha}{\bar\beta}{\bar\gamma}{\bar\delta}}\,
{{\sigma}^{\alpha}{\sigma}^{\beta}{\sigma}^{\gamma}{\sigma}^{\delta}\over%
\sigma} ,\label{eq:GABCD1}\ee  
where before expanding, a homogeneization of the indices by the parallel 
transport bivector
${{\bar g}^{\mu}}_{\bar\nu}$, i.e, $\sigma ^{\mu}={{\bar 
g}^{\mu}}_{\bar\nu}\,\sigma ^{\bar\nu},$
has been applied. The covariant expansions at $x$ are given by,
 
\bb {\bar A}=A-{1\over 2}\, A_{;\alpha}\,{\sigma}^{\alpha}+
{1\over 8}\, A_{;\alpha\beta}\,{\sigma}^{\alpha}\,{\sigma}^{\beta}+O(\epsilon 
^3),
\;\;\;\; {\bar B}=B+O(\epsilon ),\label{eq:CovAB}\ee
 
\bb {{\bar g}^{\bar\alpha}}_{\alpha}\,{{\bar 
g}^{\bar\beta}}_{\beta}\,{C}_{{\bar\alpha}{\bar\beta}}=
{C}_{{\alpha}{\beta}}+O(\epsilon ),\;\;\;\;
{{\bar g}^{\bar\alpha}}_{\alpha}\,{{\bar g}^{\bar\beta}}_{\beta}\,{{\bar 
g}^{\bar\gamma}}_{\gamma}\,{{\bar
g}^{\bar\delta}}_{\delta}\, D_{{\bar\alpha}{\bar\beta}{\bar\gamma}{\bar\delta}}=
D_{{\alpha}{\beta}{\gamma}{\delta}}+O(\epsilon ).\label{eq:CovCD}\ee
 
Now we can apply the differential operator (\ref{eq:Dopdif}) to 
(\ref{eq:GABCD1}). We will consider the two
physically relevant cases of the {\it minimal coupling} ($\xi =0$) which 
should provide a good qualitative
description for gravitons, and of the {\it conformal coupling} ($\xi =1/6$)
which should provide a good qualitative
description for photons (the use of $\xi$ as the coupling parameter
should not be confused with the coordinate $\xi$ of the interaction region). 
If we introduce the
operators,
 
\bb {\cal L}^{(1)}_{\mu\nu}=\nabla _{\mu '}\nabla _{\nu}+\nabla _{\nu '}\nabla 
_{\mu },
\;\;\;\; {\cal L}^{(2)}_{\mu\nu}=\nabla _{\mu '}\nabla _{\nu '}+\nabla _{\nu 
}\nabla _{\mu },
\label{eq:L12opdif}\ee
the operator (\ref{eq:Dopdif}), for the minimal and conformal cases, is
 
$${\cal D}^{\xi =0}_{\mu\nu}={1\over 4}\,\left[{\cal L}^{(1)}_{\mu\nu}-
{1\over 2}\, g_{\mu\nu}\, g^{\alpha\beta}\,{\cal 
L}^{(1)}_{\alpha\beta}\right],$$
 
$${\cal D}^{\xi =1/6}_{\mu\nu}={1\over 12}\,\left[2\,\left({\cal 
L}^{(1)}_{\mu\nu}-
{1\over 4}\, g_{\mu\nu}\, g^{\alpha\beta}\,{\cal L}^{(1)}_{\alpha\beta}\right)
-\left({\cal L}^{(2)}_{\mu\nu}-
{1\over 4}\, g_{\mu\nu}\, g^{\alpha\beta}\,{\cal L}^{(2)}_{\alpha\beta}\right)
\right].$$
By the properties of the geodetic interval bivector $\sigma ^{\mu}$ given in 
the Appendix B, which can be also
written as 
$\sigma ^{\mu}=2\epsilon\, p^{\mu}$ ($p^{\mu}p_{\mu}=\Sigma$) we can prove the 
following identities,
 
$$\lim _{x\rightarrow x'}{\cal L}^{(1)}_{\mu\nu}\,\sigma =
-\lim _{x\rightarrow x'}{\cal L}^{(2)}_{\mu\nu}\,\sigma =-2\, 
g_{\mu\nu},\;\;\;\; \lim _{x\rightarrow x'}{\cal
L}^{(1)}_{\mu\nu}\,{\sigma}^{\alpha}{\sigma}^{\beta}= -\lim _{x\rightarrow 
x'}{\cal
L}^{(2)}_{\mu\nu}\,{\sigma}^{\alpha}{\sigma}^{\beta}= 
-4\,\delta^{\alpha}_{(\mu}\,\delta^{\beta}_{\nu )},$$
 
\bban\lim _{x\rightarrow x'}{\cal L}^{(1)}_{\mu\nu}\,
{{\sigma}^{\alpha}{\sigma}^{\beta}{\sigma}^{\gamma}{\sigma}^{\delta}\over%
\sigma}&=&
-\lim _{x\rightarrow x'}{\cal L}^{(2)}_{\mu\nu}\,
{{\sigma}^{\alpha}{\sigma}^{\beta}{\sigma}^{\gamma}{\sigma}^{\delta}\over%
\sigma}=
\\
&&
-{48\over\Sigma}\,\delta^{(\alpha}_{(\mu}\,\delta^{\beta}_{\nu )}\, 
p^{\gamma}p^{\delta )}
+64\,\delta^{(\alpha}_{(\mu}\, p^{\beta}_{\nu )}\, p^{\gamma}p^{\delta )}
+8\, 
p^{\alpha}p^{\beta}p^{\gamma}p^{\delta}\,\left[g_{\mu\nu}-4{p_{\mu}p_{\nu}\over%
\Sigma}\right],\eean
where $(\cdots)$ is the usual symmetrization operator. Note that by a 
straightforward application of Synge's
theorem (see Appendix B), the coincidence limits of 
${\cal L}^{(1)}_{\mu\nu}$ and ${\cal L}^{(2)}_{\mu\nu}$ differ in a sign when 
they are applied over bitensors
for which the first covariant derivative has null coincidence limit. But when 
they are applied
over the biscalar $\bar A$ in (\ref{eq:CovAB}) they coincide:
$\lim _{x\rightarrow x'}{\cal L}^{(1)}_{\mu\nu}\,{\bar A}=
\lim _{x\rightarrow x'}{\cal L}^{(1)}_{\mu\nu}\,{\bar A}=A_{;\mu\nu}/2$.
The application of the opperators (\ref{eq:L12opdif})
over quartic products of $\sigma ^{\mu}$ gives  path dependent terms when the 
limit $x\rightarrow x'$
is performed. Therefore the averaging  
discussed in Appendix B is required. Provided we
work with a diagonal metric (as is the case), we only need the following 
averages,
 
$$\langle p^x p^x \rangle = {\Sigma\over 4}\, g^{xx},\;\;\;\; \langle p^x p^x 
p^x p^{\mu} \rangle ={1\over 8}\,
\left(g^{xx}\right)^2\, \delta ^{x\mu},$$
$$\langle p^x p^x p^x p^x p^{\mu} p^{\nu}\rangle
={\Sigma\over 16}\, \left(g^{xx}\right)^3\, \delta ^{x\mu}\delta 
^{x\nu}+{\Sigma\over 64}\, g^{\mu\nu}\,
\left(g^{xx}\right)^2.
$$
 
We now have all we need to evaluate the vacuum expectation value of the 
stress-energy tensor
by application of the differential operator (\ref{eq:Dopdif}) to the 
expression (\ref{eq:GABCD1}) of
$G^{(1)}_B(x,x')$. In the orthonormal basis, $\theta _0=M(1+\sin\xi)\,d\xi$,
$\theta _1=M(1+\sin\xi)\, d\eta$, $\theta _2=\cos\xi (1+\sin\xi)^{-1}\, dx$, 
$\theta _3=(1+\sin\xi)\cos\eta\, dy$, we obtain
the following  expectation values $\langle T^B_{\mu\nu}\rangle$ for the 
minimal and conformal couplings
 
\bb\langle T^B_{\mu\nu}\rangle ^{\xi =0} =
\lim _{x\rightarrow x'}{\cal D}^{\xi =0}_{\mu\nu}\, G^{(1)}_B(x,x')=
{\rm diag}\left(\rho _1 ,\, -\rho _1 ,\, \rho _2 ,\, -\rho 
_1\right),\label{eq:TMNminim}
\ee
 
\bb\langle T^B_{\mu\nu}\rangle ^{\xi =1/6} =
\lim _{x\rightarrow x'}{\cal D}^{\xi =1/6}_{\mu\nu}\, G^{(1)}_B(x,x')={\rm 
diag}\left(\rho ,\, -\rho ,\, 3\rho
,\, -\rho\right),\label{eq:TMNconforme}\ee
where $\rho$, $\rho _1$ and $\rho _2$ are positive definite functions given by,
 
$$\rho _1 ={\cos ^{-4}\xi \over 256\pi ^2\, M^4}\,{10771\over 
2880}\left(1+\lambda _1\,\cos
^2\xi +O(\cos ^4\xi )\right),$$
$$\rho _2 ={\cos ^{-4}\xi \over 256\pi ^2\, M^4}\,{3341\over
9216}\left(1+\lambda _2\,\cos ^2\xi +O(\cos ^4\xi )\right),$$
$$\rho ={\cos ^{-4}\xi \over 256\pi ^2\, M^4}\,{ 6869\over 
2880}\left(1+\lambda \,\cos ^2\xi
+O(\cos ^4\xi )\right),$$ 
which are unbounded at the horizon ($\xi =\pi /2$), and the approximate values 
for $\lambda _1$,
$\lambda _2$ and $\lambda$ are
$\lambda _1=1.474\cdots$, $\lambda _2=1.466\cdots$ and $\lambda =1.469\cdots$. 
From (\ref{eq:GBT})
this is not the final result of the stress-energy tensor in region I, $\langle 
T_{\mu\nu}\rangle _I$, since the
trace anomaly term  must be included. But this term depends on the spacetime 
curvature which is finite at the
horizon, 
 
$${a_2\over 64\,\pi ^2}\, g_{\mu\nu}=\left[{1\over 240\, M^4\,\pi ^2}\,{1\over 
64}+O(\cos ^2\xi)\right]\,
{\rm diag}\left(1 ,\, -1 ,\, -1,\, -1\right),$$ 
and it does not modify the dominant contributions of (\ref{eq:TMNminim})
or (\ref{eq:TMNconforme}), which increase as $\cos ^{-4}\xi$ near the horizon. 
In the conformally coupled case
the trace $\langle T^{\mu}_{\mu}\rangle _I$ is finite and can be obtained from 
the previous term.
Inspection of (\ref{eq:TMNminim})-(\ref{eq:TMNconforme}) shows that for both 
couplings the weak energy
condition is satisfied \cite{wal84}, which means that the energy density is 
nonnegative for any observer, and
that the strong energy condition is only satisfied for the conformal coupling.

\section{Conclusions}

We have evaluated the expectation value of the stress-energy tensor of a 
massless scalar field in
the state, which corresponds to the
physical vacuum state before the collision of the plane waves. This vacuum 
state has an analogue in the
Schwarzchild black hole case as the empty state at large radius from the hole, 
the Boulware vacuum
state, for which the vacuum expectation value of the stress-energy tensor also 
diverges at the horizon. In
this case, however, it is argued that the Boulware vacuum state is not 
physical since it does not
correspond to the vacuum of the gravitational collapse problem 
\cite{fro85tho86,can80sci81}.
 
Our results are the following. Before the collision of the plane waves (in 
region IV) 
$\langle T^{\mu\nu}\rangle _{IV}=0$ by the definition of our vacuum state, in 
the plane wave regions (regions
II and III) we have found in section 7 that $\langle T^{\mu\nu}\rangle 
_{II}=\langle T^{\mu\nu}\rangle
_{III}=0$, since the plane waves do not polarize the vacuum, and finally, in 
the interaction region (region I)
$\langle T^{\mu\nu}\rangle _{I}$ becomes unbounded at the regular 
Killing-Cauchy horizon
($\xi =\pi /2$) in the conformally and minimally coupled cases. In both cases 
the 
weak energy condition is satisfied, the rest energy density is positive and 
diverges as $\cos ^{-4}\xi$ and
two of the principal pressures are negative and of the same order of magnitude 
of the energy density. The
strong energy condition is satisfied for the conformal coupling, in this case 
$\langle T^{\mu}_{\mu}\rangle _I$ is finite but  $\langle
T^{\mu\nu}\rangle \langle T_{\mu\nu}\rangle _I$ diverges at the horizon and we 
may
use ref \cite{hel93} on the stability of Cauchy horizons to argue that the 
horizon will aquire by backreaction
a curvature singularity too.
Thus contrary to the simple plane waves
which do not polarize the vacuum \cite{gib75,des75}, the nonlinear collision 
of these waves polarize the vacuum
and the focusing effect that the waves exert produce at the focusing
points an unbounded positive energy density.
Therefore when the colliding waves produce a Killing-Cauchy horizon that 
horizon is unstable by vacuum
polarization.
 
In the more generic case when the wave collision produces a spacelike 
singularity it seems clear that the
vacuum expectation value of the stress-energy tensor will also grow unbounded 
near the singularity. The
reason is that such unboundness is, essentially, a consequence of the 
blueshift suffered by the mode solutions
as they enter the plane wave regions, and it is easy to see \cite{gar91} that 
any plane wave produces a
similar blueshift on mode solutions. In view of our results it seems very 
unlikely that the negative pressures
associated to the quantum fields could prevent the formation of the singularity.

\vskip 1.25 truecm
 
{\Large{\bf Acknowledgements}}
 
\vskip 0.5 truecm
 
\noindent
We are grateful to E. Calzetta, A Campos, D. Espriu, A. Feinstein, J. Garriga 
and J. Iba{\~n}ez for helpful
discussions. This work has been partially supported by  CICYT Research 
Projects No. AEN95-0882 and
AEN95-0590, and the European Project, No. CI1-CT94-0004.

\appendix

\section{Adiabatic expansion of ``in" modes in region I}
 
With the parameters of section 5, defining $p\equiv Lq^{-1}(\cos\xi )^{-1}$
and $p_1\equiv Lq^{-1}(\cos\eta )^{-1}$ and assuming that $p\ll 1$, $p_1\ll 1$,
we can expand (\ref{eq:phi}) up to adiabatic order 4 in powers of $p$ and 
$p_1$. The process requires to expand
the WKB functions (\ref{eq:egam}) and (\ref{eq:varphiW}). These expansions are 
rather easy to perform and, in
particular, the expansion for (\ref{eq:varphiW}) allow us to fix the two 
arbitrary coefficients
$C_1$ and $C_2$ by requiring that (\ref{eq:varphiW}) corresponds asymptotically
to the function (\ref{eq:philegendre}) which is, modulus a constant factor, 
an associated Legendre function. Using that the asymptotic expansion
of an associated Legendre function $P_l^m(\sin\eta)$, in the range of values,
$Lm(\cos\eta )^{-1}\ll 1$, is \cite{gra80}, 
 
$$ P_l^m(\sin\eta)=l^m\left\{{l^{-1/2}\over\sqrt{\pi\cos\eta}}\,
\cos\left[(l+{1\over 2})\eta -(l+m){\pi\over 2}\right]+O(l^{-3/2})\right\}, $$
we find that
 
$$ \varphi (\eta)={\sqrt{2l+1}\over 4\pi\sqrt{M}}\,{1\over\sqrt{W_2^{{\rm 
(4)}}(\eta)}}\,
{\rm e}^{i(l+m)\pi /2}
\left\{{\rm e}^{-i\int { W}_2^{{\rm (4)}}(\eta)\, d\eta ^*}+
(-1)^{l+m}\,{\rm e}^{i\int{ W}_2^{{\rm (4)}}(\eta)\, d\eta ^*}\right\}.$$
 
With these results, we can divide each term (\ref{eq:AA}) or (\ref{eq:BA}) in 
four integrals such that the
variation of the phase of their integrand functions are given by
 
$${{\dot A}(u)\over 2L_2k_-} \pm {W}^{\rm(4)}_-(u),\;\;\;\;{{\dot A}(u)\over 
2L_2k_-} \pm {W}^{\rm(4)}_+(u),$$
$${{\dot A}(u)\over 2L_1k_+} \pm {W}^{\rm(4)}_-(u),\;\;\;\;{{\dot A}(u)\over 
2L_1k_+} \pm {W}^{\rm(4)}_+(u).$$
and the modulus of their integrand functions are given, respectively, by
 
$$F_1(u)\,\left({{\dot A}(u)\over 2L_2k_-} \mp 
{W}^{\rm(4)}_-(u)\right)+F_2(u),\;\;\;\;
F_1(u)\,\left({{\dot A}(u)\over 2L_2k_-} \mp {W}^{\rm(4)}_+(u)\right)+F_2(u),$$
$$F_1(u)\,\left({{\dot A}(u)\over 2L_1k_+} \mp 
{W}^{\rm(4)}_-(u)\right)+F_2(u),\;\;\;\;
F_1(u)\,\left({{\dot A}(u)\over 2L_1k_+} \mp {W}^{\rm(4)}_+(u)\right)+F_2(u).$$
Note that when a sign $\pm$ is taken in a phase term, then in the modulus term 
we have
the oposite sign $\mp$. 
Up to adiabatic order 4 all these functions, i.e. ${W}^{\rm(4)}_\pm (u)$, 
$F_1(u)$, $F_2(u)$ and 
${\dot A}(u)$, can be expanded in powers of
$p$ on the boundaries between regions II and III with region I, where $\xi 
=\pm\eta$ and thus $p=p_1$, and 
where we use for simplicity $z\equiv\sin u$. We get,
 
\bban {W}^{\rm(4)}_+(u)&=&L^{-1}\left\{
2+
p^2\left(
{q^2\over 4}\, (2-z^2)
+{V^2\over 32}\,(1+z)^4-{U^2\over 2}\right)+\right. \\
& &
\left. 
+p^4\left[
-{q^4\over 64}\, (12+12z^2+z^4)
+q^2\,\left(-{V^2\over 256}\,(1+z)^4(30-24z+7z^2)+\right.\right.\right.\\
& &
\left.\left.\left.{U^2\over 16}\,(6+7z^2)\right)
 -{V^4\over 2048}\,(1+z)^8-{U^4\over 8}
\right] 
\right\} +O(L^{-1}p^6), \\
& &\\
{W}^{\rm(4)}_-(u)&=&L^{-1}\left\{
p^2\left(
{V^2\over 32}\,(1+z)^4+{U^2\over 2}\right)+\right. \\
& &
\left. 
+p^4\left[
q^2\,\left(-{V^2\over 256}\,(1+z)^4(30-24z+7z^2)-{U^2\over 
16}\,(6+7z^2)\right)-\right.\right.\\
& &
\left.\left. -{V^4\over 2048}\,(1+z)^8+{U^4\over 8}
\right] 
\right\} +O(L^{-1}p^6),   \\
& &\\
{F}_1(u)&=&L\, M\left\{
2+
p^2\left(
-{q^2\over 4}\, (2-z^2)
-{V^2\over 32}\, (1+z)^4+{U^2\over 2}\right)+\right. \\
& &
\left. 
+p^4\left[
{q^4\over 64}\, (20+4z^2+3z^4)
+q^2\,\left({V^2\over 256}\,(1+z)^4(34-24z+5z^2)-\right.\right.\right.\\
& &
\left.\left.\left.{5\, U^2\over 16}\,(2+z^2)\right) +{5\, V^4\over 
4096}\,(1+z)^8+{5\, U^4\over 16}
-{V^2\, U^4\over 128}\,(1+z)^4
\right] 
\right\} +O(Lp^6),   \\
& &\\
{F}_2(u)&=&L\, p^3\, M\left\{
-{q^2\over 2}\, z
-{V^2\over 16}\, (1+z)^3(2+z-z^2)+{U^2}\, z\right\}+O(Lp^5). \\
& &\\
{\dot A}(u)&=&L^{-2}\, p^2\left\{
{V^2\over 16}\,(1+z)^4+{U^2}\right\}. \eean

Among the integrals in (\ref{eq:AA})-(\ref{eq:BA}), there is only one integral 
which contains stationary points
and it is always in the term
${\cal A}^{\rm (A)}_{kk'}$ in (\ref{eq:AA}) when either $L^{-1}=L_2k_-$ or 
$L^{-1}=L_1k_+$, because the
variation of the phase takes in this case its minimum value with respect to 
the modulus, and its contribution
becomes dominant; consequently we can neglect the remaining integrals. The 
dominant integrals
connecting the ``in" modes between regions II and I can be written as,
 
$$\int du\, F_1(u)\,\left({{\dot A}(u)\over 2L_2k_-} + 
{W}^{\rm(4)}_-(u)\right)\,
{\rm e}^{\int du \left({{\dot A}(u)/ (2L_2k_-)} - {W}^{\rm(4)}_-(u)\right)},$$
which can be written in powers of $p$ as (\ref{eq:intdominant}).
The dominant integrals connecting the ``in" modes between regions III and I 
can be written in the same way but
changing $L_2k_-$ for $L_1k_+$.

\section{Bitensor algebra}
 
A {\it bitensor} with $n+m$ indices, $\alpha _1,\alpha _2,\cdots ,\alpha _n$,
$\beta _1,\beta _2,\cdots ,\beta _m$,
$T_{\alpha _1,\alpha _2,\cdots ,\alpha _n\beta _1,\beta _2,\cdots ,\beta _m}$,
is an object which transforms,
under coordinate changes, as the product of two ordinary tensors, $A_{\alpha 
_1\alpha _2\cdots \alpha _n}$ and
$B_{\beta '_1\beta '_2\cdots \beta '_m}$, each at a different spacetime point 
\cite{dew60,chr76,chr78}.
The standard  tensorial operations can be straightforwardly extended to 
bitensors with the simple precaution of
contracting only indices refering to the same spacetime point, and noting that 
when a covariant derivative is
performed over one of the points, the indices referring to the other point are 
irrelevant.
 
If we have an arbitrary bitensor evaluated at two close spacetime points $x$ 
and $x'$ it is possible to
expand it in a covariant way at one of the two points. To achieve this we make 
use of the {\it geodetic
interval}, $s(x,x')$, a basic object in geodesic theory, which describes the 
proper
distance between the spacetime points $x$ and $x'$ along a non null geodesic 
connecting them. Let us restrict 
ourselves to a neighbourhood of $x$ (or $x'$) where the geodesics emanating 
from $x$ do not cross (mathematicaly
this means that the unique geodesic
joining
$x$ and
$x'$ is an extremal curve \cite{wal84}), then $s(x,x')$ is single-valued. The 
basic properties of
$s(x,x')$ are 
 
\bb g^{\mu \nu }s_{,\mu }s_{,\nu }=g^{\mu '\nu '}s_{,\mu '}s_{,\nu '}=\pm 
1,\label{eq:s1'}\ee
together with the symmetry $s(x,x')=s(x',x)$ and $\lim _{x\rightarrow 
x'}s(x,x')=0$.
 
These ensure that the bitensor
$g^{\mu \nu }(x)s_{,\mu }(x,x')$ is a tangent vector to the geodesic at the 
point $x$ and a scalar at $x'$.
Note that
$g^{\mu \nu }(x)s_{,\mu }(x,x')$ is an orthogonal vector to the surface 
$s(x,x')={\rm constant}$, and this
surface is orthogonal to the geodesics which emanate from  $x'$ and therefore
$g^{\mu \nu }(x)s_{,\mu }(x,x')$ is a tangent vector to the geodesic that 
crosses $s(x,x')={\rm constant}$ at
the point $x$. Of course, if $x$ is fixed and we draw the geometrical locus of 
the points $x'$
at the same proper distance $s(x',x)$, from $x$ an equivalent description is 
found. If we take the signature
$(+,-,-,-)$ then the positive sign in (\ref{eq:s1'}) refers to timelike 
separated points, and the negative to 
spacelike separated points. When
$s(x,x')=0$, then $x$ and $x'$ lie on a light cone. 
 
Instead of $s(x,x')$ it is convenient to define the {\it geodetic biscalar}, 
$\sigma (x,x')$, which unlike
$s(x,x')$ has no branchpoints, as
 
\bb \sigma ={s^2\over 2}\,\Sigma ,\label{eq:si}\ee
where we introduce the notation $\Sigma =\pm$. For the signature 
$(+,-,-,-)$ the plus sign is for timelike separated points and the negative 
sign for spacelike
separated points. Let us define $\sigma _{\mu}\equiv \sigma _{,\mu}$,
then the following relations are trivially satisfied,
 
\bb {1\over 2}\, g^{\mu \nu }\sigma _{\mu }\sigma _{\nu }=
{1\over 2}\, g^{\mu '\nu '}\sigma _{\mu '}\sigma _{\nu '}=\sigma 
,\label{eq:si0}\ee
together with the coincidence limits
$\lim _{x\rightarrow x'}\sigma=0$, and
$\lim _{x\rightarrow x'}\sigma _{,\mu}=\lim _{x\rightarrow x'}\sigma _{,\mu 
'}=0$.
 
Given an arbitrary bitensor $A_{\mu\nu}(x,x')$, with both indices referring to 
the same point $x$, it is
possible to  covariantly expand this tensor in the neighbourhood of $x$, as
 
\bb 
A_{\alpha\beta}(x,x')=A_{\alpha\beta}(x)+{A^{\gamma}}_{\alpha\beta}(x)\,%
\sigma_{\gamma}+{1\over
2}\, {A^{\gamma\delta}}_{\alpha\beta}(x)\sigma_{\gamma}\sigma_{\delta}+\cdots 
,\label{eq:Tc}\ee
where the coefficients
$A_{\alpha\beta}(x)$, ${A^{\gamma}}_{\alpha\beta}(x),\cdots$ are ordinary 
local tensors at the point $x$. These
coefficients are easily calculated, provided that the successive covariant 
derivatives of the geodetic biscalar
$\sigma$ are known. For this we derive (\ref{eq:si0}) recursively:
 
\bb \sigma _{,\mu}=g^{\alpha \beta }\sigma _{\alpha }\sigma _{;\beta\mu }, 
\;\;\;\; \sigma _{;\mu\nu}=g^{\alpha
\beta }\left(
\sigma _{;\alpha\nu }\sigma _{;\beta\mu }+\sigma _{\alpha }\sigma 
_{;\beta\mu\nu }\right),  
\label{eq:s12}\ee
\bb \sigma _{;\mu\nu\tau}=g^{\alpha \beta }\left(
\sigma _{;\alpha\nu\tau }\sigma _{;\beta\mu }+\sigma _{;\alpha\nu }\sigma 
_{;\beta\mu\tau }
+\sigma _{;\alpha\tau}\sigma _{;\beta\mu\nu}+\sigma _{\alpha }\sigma 
_{;\beta\mu\nu\tau}\right).
 \label{eq:s3}\ee
Then it is straightforward to find,
 
$$A_{\alpha\beta}=\left[A_{\alpha\beta}\right],\;\;\;\; 
A_{\alpha\beta\gamma}=\left[A_{\alpha\beta
;\gamma}\right]-A_{\alpha\beta ;\gamma},$$
$$A_{\alpha\beta\gamma\delta}=\left[A_{\alpha\beta ;\gamma\delta}\right]-
A_{\alpha\beta ;\gamma\delta}
-A_{\alpha\beta\gamma ;\delta}-A_{\alpha\beta\delta ;\gamma},$$
where the notation introduced by Synge to indicate the coincidence limit 
$x\rightarrow x'$
is used, i.e,
$\left[T_{\alpha _1,\alpha _1,\cdots ,\alpha _n\beta _1,\beta _1,\cdots ,\beta 
_m}\right]\equiv
\lim _{x\rightarrow x'}T_{\alpha _1,\alpha _1,\cdots ,\alpha _n\beta _1,\beta 
_1,\cdots ,\beta _m}(x,x')$.
In particular, it is possible to covariantly expand the successive covariant 
derivatives of the geodetic
biscalar $\sigma$ as,
 
$$
\sigma _{;\alpha\beta}=g_{\alpha\beta}+{1\over
3}\,{{{R_{\alpha}}^{\mu}}_{\beta}}^{\nu}\sigma_{\mu}\sigma_{\nu}+O(s^3),\;\;\;\;
\sigma _{;\alpha\beta\gamma}={1\over
3}\left({R_{\alpha\gamma\beta}}^{\mu}+{{R_{\alpha}}^{\mu}}_{\beta\gamma}%
\right)\sigma_{\mu}+O(s^2),$$
$$
\sigma _{;\alpha\beta\gamma\delta}={1\over
3}\left(R_{\alpha\gamma\beta\delta}+R_{\alpha\delta\beta\gamma}\right)+O(s). 
$$
 
Sometimes, for the sake of symmetry, it is convenient to  covariantly expand 
the bitensor in terms of the
midpoint
$\bar x$  of the geodesic connecting $x$ and $x'$. For simplicity let us take 
a biscalar $G(x,x')$. Then
a midpoint covariant expansion reads,
 
$$G(x,x')=A({\bar x})+{A_{\mu}}({\bar x})\,{\bar\sigma}^{\mu}+
{A_{(\mu\nu )}}({\bar x})\,{\bar\sigma}^{\mu}\,{\bar\sigma}^{\nu}+
{A_{(\mu\nu\tau )}}({\bar 
x})\,{\bar\sigma}^{\mu}\,{\bar\sigma}^{\nu}\,{\bar\sigma}^{\tau}
+\cdots ,$$
where $(\cdots )$ index symmetrization.
The relation between the expansion coefficients $A({\bar x})$ and the biscalar 
$G(x,x')$ is easily
found to be,
 
$$A=\left[G\right],\;\;\;\; A_{\mu}=\left[G_{;\mu}\right]-{1\over 2}\, 
\left[G\right]_{;\mu},\;\;\;\; 2!\,
A_{\mu\nu}{\dot =}\left[G_{;\mu\nu}\right]-\left[G_{;\mu}\right]_{;\nu} 
+{1\over 4}\,
\left[G\right]_{;\mu\nu},$$
$$3!\, A_{\mu\nu\tau}{\dot =}\left[G_{;\mu\nu\tau}\right]-{3\over 
2}\,\left[G_{;\mu\nu}\right]_{;\tau}
+{3\over 4}\, \left[G_{;\mu}\right]_{;\nu\tau}-{1\over 8}\, 
\left[G\right]_{;\mu\nu\tau},$$
where the notation $\dot =$ means that the equality is modulus terms which 
vanish under symmetrizations.
 
The covariant expansion of a bitensor with  indices referring to different 
points is slightly more
sophisticated because it is necessary to homogenize the indices before the 
expansion. To do this we use the
{\it parallel transport bivector}, ${{\bar g}^{\mu}}_{\mu '}(x,x')$, which is 
defined by analogy to the parallel
transport of ordinary vectors on a manifold. Given an ordinary vector ${ 
A}^{\mu}(x)$ at the point
$x$ it is possible to parallel transport it at the point $x'$ on a non-null 
geodesic, say ${ A}^{\mu '}(x')$,
by a linear operator
${{\bar g}^{\mu}}_{\mu '}(x,x')$ as
${ A}^{\mu '}(x')= {{\bar g}_{\mu}}^{\mu '}(x,x')\, { A}^{\mu
}(x)$,
provided that ${{\bar g}^{\mu}}_{\mu '}(x,x')$ is covariantly conserved along 
the geodesic, i.e.
${\sigma}^{\nu '}(x,x')\, {{\bar g}_{\mu}}^{\mu '}(x,x')_{;\nu '}=0$,
and it satisfies the trivial boundary condition,
$\left[{{\bar g}_{\mu}}^{\mu '}\right]={\delta}^{\mu '}_{\mu }$.
In exactly the same way, we can define a parallel transport for bitensors as 
${{\bar g}^{\mu}}_{\mu
'}(x,x')$ if it verifies the previous differential condition not only at the 
point $x'$ but also
at the point $x$, i.e,
 
\bb{\sigma}^{\nu '}(x,x')\,{{\bar g}_{\mu}}^{\mu '}(x,x')_{;\nu '}=0,\;\;\;\; 
{\sigma}^{\nu }(x,x')\,{{\bar
g}_{\mu }}^{\mu '}(x,x')_{;\nu }=0,      \label{eq:gtpxx'}\ee  
and the same boundary condition $\left[{{\bar g}_{\mu}}^{\mu 
'}\right]={\delta}^{\mu '}_{\mu }$.
The parallel transport bivector, defined in this a way, is unique because if 
we fix the point $x$ one can
integrate the second equation in (\ref{eq:gtpxx'}) along each geodesic which 
emanates from $x$  up to $x'$ with
the initial condition 
$\left[{{\bar g}_{\mu}}^{\mu '}\right]={\delta}^{\mu '}_{\mu }$. Reciprocally, 
if we fix $x'$, the first
equation in (\ref{eq:gtpxx'}) can be integrated from $x'$ down to $x$  with 
the same initial condition. This
reciprocity can be expressed as,
${{\bar g}^{\mu}}_{\mu '}(x,x')={{\bar g}_{\mu '}}^{\mu }(x',x)$.
From the geometric interpretation of ${{\bar g}^{\mu}}_{\mu '}(x,x')$ it is 
straightforward to see that:
 
$${{\bar g}_{\mu}}^{\mu '}\,{{\bar g}^{\nu }}_{\nu '}\, g_{\mu '\nu '}=
g_{\mu\nu},\;\;\;\;{{\bar g}^{\mu }}_{\mu '}\,{{\bar g}^{\nu }}_{\nu '}\, 
g_{\mu \nu }=
g_{\mu '\nu '},\;\;\;\;{{\bar g}_{\mu}}^{\mu '}\,\sigma _{\mu '}=-\sigma _{\mu 
},$$
 
$${{\bar g}^{\mu }}_{\mu '}\,\sigma _{\mu }=-\sigma _{\mu '},\;\;\;\;{{\bar 
g}_{\mu}}^{\mu '}\,{{\bar g}^{\nu
}}_{\mu '}=\delta ^{\nu}_{\mu},
\;\;\;\;{{\bar g}_{\mu}}^{\mu '}\,{{\bar g}^{\mu }}_{\nu '}=\delta ^{\mu 
'}_{\nu '}.
$$
 
Thus given a non-homogeneous bitensor $A_{\mu\nu '}(x,x')$, we first 
homogenize it as,
${\hat A}_{\mu\nu}(x,x')={{\bar g}_{\nu }}^{\nu '}(x,x'){A}_{\mu\nu '}(x,x')$,
and then we covariantly expand ${\hat A}_{\mu\nu}(x,x')$ in the same way as
(\ref{eq:Tc}).
 
An interesting result in the algebra of bitensors, usually known as Synge's 
theorem \cite{syn73}, is that given
an arbitrary bitensor $T_{\alpha _1\alpha _2\cdots \alpha _n\beta '_1\beta 
'_2\cdots \beta '_m}$, whose 
coincidence limit and the coincidence limit of its first covariant derivatives 
exist, then the following
identity is satisfied
\cite{chr76}:
 
\bb \left[T_{\alpha _1\alpha _2\cdots \alpha _n\beta '_1\beta '_2\cdots \beta
'_m;\mu '}\right]=-\left[T_{\alpha _1\alpha _2\cdots \alpha _n\beta '_1\beta 
'_2\cdots \beta
'_m;\mu}\right]+\left[T_{\alpha _1\alpha _2\cdots \alpha _n\beta '_1\beta 
'_2\cdots \beta
'_m}\right]_{;\mu}.   \label{eq:Syn}\ee
This is very useful when the calculations involve a lot of primed indices and 
we know the coincidence
limits of the nonprimed quantities. A trivial application of this result 
implies that
$\left[\sigma_{\mu\mu '}\right]=-g_{\mu\mu '}$.
 
When one is performing the covariant expansion of Green functions in 
preparation for the calculation of
the regularized expectation value of the stress-energy tensor, ambiguities in
the coincidence limits appear when one of the points, say $x$, is held fixed 
and an arbitrary point $x'$ is
allowed to approach $x$. These ambiguities are due to the different paths that 
$x'$ may follow, therefore
some type of averaging is required. The most elementary averaging is called 
{\it four dimensional
hyperspherical averaging} \cite{adl77}  and it
consists in giving the same weight to
all the geodesics which emanate from $x$ as follows. First one analytically
continues to an Euclidean metric the components of the tangent vectors to the 
geodesics which emanate from
$x$. Second, one averages over a 4-sphere, and third the results are continued 
back to the original
metric. It is not very complicated to find the following averaging formulae 
which are useful in this paper, 
 
$$ \langle {\sigma}_{\mu}{\sigma}_{\nu} \rangle ={1\over 
2}\,\sigma\,g_{\mu\nu},\;\;\;\; \langle
{\sigma}_{\mu}{\sigma}_{\nu} {\sigma}_{\tau}{\sigma}_{\rho} \rangle = {1\over 
2}\,\sigma ^2\,
g_{\mu (\nu} g_{\tau\rho )}, \;\;\;\;
 \langle {\sigma}_{\mu}{\sigma}_{\nu}
{\sigma}_{\tau}{\sigma}_{\rho}{\sigma}_{\xi}{\sigma}_{\eta}
\rangle = {5\over 8}\,\sigma ^3\,  
g_{\mu (\nu}g_{\tau (\rho}g_{\xi\eta ))}
$$
Note that from the symmetry of the averaging procedure, the average of an odd 
number of components $\sigma
^{\mu}$ vanishes identically.

\section{Useful  adiabatic expansions}
 
The adiabatic coefficients ${\cal A}_n$ in formula (\ref{eq:G(1)3}) are given 
by,
 
\bban {\cal A}_0&=&
{1\over\omega}
+T^2\,\left(  -{5\over 32}\,{{\dot V}_l^2\over \omega ^7} 
+ {1\over 8}\,{{\ddot V}_l\over \omega ^5}         
\right) \\
& &
+T^4\,\left(  
{1155\over 2048}\,{{\dot V}_l^4\over \omega ^{13}} 
- {231\over 256}\,{{\dot V}_l^2{\ddot V}_l\over \omega ^{11}} 
+{21\over 128}\,{{\ddot V}_l^2\over \omega ^9} 
+ {7\over 32}\,{{\dot V}_l{\dddotV}_l\over \omega ^9}  
-{1\over 32}\,{{\ddddotV}_l\over \omega ^7}        
\right), 
\\
& &
\\
{\cal A}_1&=&
T^2\,\left(  
-{5\over 16}\,{{\dot V}_l^2\over \omega ^6} 
+ {1\over 4}\,{{\ddot V}_l\over \omega ^5} 
\right)\,\xi _1^*
\\
&&
+T^4\,\left(  
{1155\over 1024}\,{{\dot V}_l^4\over \omega ^{12}} 
- {231\over 128}\,{{\dot V}_l^2{\ddot V}_l\over \omega ^{11}}
+{21\over 64}\,{{\ddot V}_l^2\over \omega ^8} 
+ {7\over 16}\,{{\dot V}_l{\dddotV}_l\over \omega ^8} 
-{1\over 16}\,{{\ddddotV}_l\over \omega ^6}        
\right)\,\xi _1^*,
\\
&&
\\
{\cal A}_2&=&
T\,\left(
-{1\over 4}\,{{\dot V}_l\over \omega ^3} 
\right)\,\xi _2^*
+T^2\,\left(  
{1\over 4}\,{{\dot V}_l^2\over \omega ^5} 
- {1\over 4}\,{{\ddot V}_l\over \omega ^3} 
\right)\,\xi _1^{*2}
\\
&&
+T^3\,\left( 
{35\over 128}\,{{\dot V}_l^3\over \omega ^9} 
- {5\over 16}\,{{\dot V}_l{\ddot V}_l\over \omega ^7}
+{1\over 16}\,{{\dddotV}_l\over \omega ^5}     
\right)\,\xi _2^*
\\
&&
+T^4\,\left( 
-{525\over 512}\,{{\dot V}_l^4\over \omega ^{11}} 
+ {217\over 128}\,{{\dot V}_l^2{\ddot V}_l\over \omega ^9}
-{11\over 32}\,{{\ddot V}_l^2\over \omega ^7} 
- {13\over 32}\,{{\dot V}_l{\dddotV}_l\over \omega ^7} 
+{1\over 16}\,{{\ddddotV}_l\over \omega ^5}        
\right)\,\xi _1^{*2},
\\
&&
\\
{\cal A}_3&=&
T\,\left(
-{1\over 2}\,{{\dot V}_l\over \omega ^2} 
\right)\,\xi _1^*\,\xi _2^*
+T^2\,\left[
\left(  
-{5\over 96}\,{{\dot V}_l^2\over \omega ^6} 
+{1\over 24}\,{{\ddot V}_l\over \omega ^2}
\right)\,\xi _3^{*}+
\left(
{1\over 12}\,{{\dot V}_l^2\over \omega ^4} 
- {1\over 6}\,{{\ddot V}_l\over \omega ^2}
\right)\,\xi _1^{*3} 
\right]
\\
&&
+T^3\,\left(  
{35\over 64}\,{{\dot V}_l^3\over \omega ^8} 
- {5\over 8}\,{{\dot V}_l{\ddot V}_l\over \omega ^6}
+{1\over 8}\,{{\dddotV}_l\over \omega ^4}     
\right)\,\xi _1^*\,\xi _2^*
\\
&&
+T^4\,\left[
\left(  
{285\over 2048}\,{{\dot V}_l^4\over \omega ^{12}} 
- {77\over 256}\,{{\dot V}_l^2{\ddot V}_l\over \omega ^{10}}
+{7\over 128}\,{{\ddot V}_l^2\over \omega ^8} 
+ {7\over 96}\,{{\dot V}_l{\dddotV}_l\over \omega ^8} 
-{1\over 96}\,{{\ddddotV}_l\over \omega ^6}
\right)\,\xi _3^{*}
\right.
\\
&&
\left.
+\left(  
-{35\over 64}\,{{\dot V}_l^4\over \omega ^{10}} 
+ {63\over 64}\,{{\dot V}_l^2{\ddot V}_l\over \omega ^8}
-{1\over 4}\,{{\ddot V}_l^2\over \omega ^6} 
- {11\over 48}\,{{\dot V}_l{\dddotV}_l\over \omega ^6} 
+{1\over 24}\,{{\ddddotV}_l\over \omega ^4}
\right)\,\xi _1^{*3}         
\right],
\\
&&
\\
{\cal A}_4&=&
T\,\left(
-{1\over 48}\,{{\dot V}_l\over \omega ^3} 
\right)\,\xi _4^*
+T^2\,\left[
\left(  
{3\over 32}\,{{\dot V}_l^2\over \omega ^5} 
-{1\over 16}\,{{\ddot V}_l\over \omega ^3}
\right)\,\xi _2^{*2}+
\left(
{1\over 12}\,{{\dot V}_l^2\over \omega ^5} 
- {1\over 12}\,{{\ddot V}_l\over \omega ^3}
\right)\,\xi _1^{*}\,\xi _3^{*} 
\right]
\\
&&
+T^3\,\left[
\left(  
{35\over 1536}\,{{\dot V}_l^3\over \omega ^9} 
- {5\over 192}\,{{\dot V}_l{\ddot V}_l\over \omega ^7}
+{1\over 192}\,{{\dddotV}_l\over \omega ^5}
\right)\,\xi _4^*
+\left(  
-{15\over 32}\,{{\dot V}_l^3\over \omega ^7} 
+ {9\over 16}\,{{\dot V}_l{\ddot V}_l\over \omega ^5}
-{1\over 8}\,{{\dddotV}_l\over \omega ^3}
\right)\,\xi _1^{*2}\,\xi _2^*    
\right]
\\
&&
+T^4\,\left[
\left(  
-{315\over 1024}\,{{\dot V}_l^4\over \omega ^{11}} 
+ {245\over 512}\,{{\dot V}_l^2{\ddot V}_l\over \omega ^9}
-{5\over 64}\,{{\ddot V}_l^2\over \omega ^7} 
- {15\over 128}\,{{\dot V}_l{\dddotV}_l\over \omega ^7} 
+{1\over 64}\,{{\ddddotV}_l\over \omega ^5}
\right)\,\xi _2^{*2}
\right.
\\
&&
\left.
+\left(  
-{175\over 512}\,{{\dot V}_l^4\over \omega ^{11}} 
+{217\over 384}\,{{\dot V}_l^2{\ddot V}_l\over \omega ^9}
-{11\over 96}\,{{\ddot V}_l^2\over \omega ^7} 
- {13\over 96}\,{{\dot V}_l{\dddotV}_l\over \omega ^7} 
+{1\over 48}\,{{\ddddotV}_l\over \omega ^5}
\right)\,\xi _1^{*}\,\xi _3^{*}
\right.         
\\
&&
\left.
+\left(  
{35\over 192}\,{{\dot V}_l^4\over \omega ^9} 
- {37\over 96}\,{{\dot V}_l^2{\ddot V}_l\over \omega ^7}
+{13\over 96}\,{{\ddot V}_l^2\over \omega ^5} 
+ {1\over 12}\,{{\dot V}_l{\dddotV}_l\over \omega ^5} 
-{1\over 48}\,{{\ddddotV}_l\over \omega ^3}
\right)\,\xi _1^{*4}         
\right],
\\
&&
\\
 {\cal A}_5&=&
T\,\left(
-{1\over 12}\,{{\dot V}_l\over \omega ^2}\,\xi _2^*\,\xi _3^* 
-{1\over 24}\,{{\dot V}_l\over \omega ^2}\,\xi _1^*\,\xi _4^* 
\right) +O(T^2) , \eean
 
The coefficients for the Hadamard function in (\ref{eq:G(1)fR}), using
$z=\sin\xi$ are:
 
\bban 16M^2\pi ^2{\bar
A}&=&(19136+1523z+99z^2-950z^3-422z^4+3z^5+3z^6){(1+z)^{-3}\over 14400},
\\
& &
\\
16M^4\pi ^2{\bar B}&=&(-1093271+1317560z-275260z^2-187012z^3+37110z^4-
\\
& &
4908z^5-2396z^6+
772z^7+57z^8-12z^9)
(1+z)^{-6}{(1-z^2)^{-2}\over 115200}+
\\
& &
2(4z-3)(1+z)^{-6}(1-z)^{-1}\,\ln\left[(1+z)\,
2^{-1/3}\right],
\\
& &
\\
16M^4\pi ^2C_{{\bar x}{\bar x}}&=&
(-1474763+1403169z-366148z^2-39568z^3+18686z^4-
774z^5-
\\
& &
196z^6+
56z^7+21z^8-3z^9)
(1+z)^{-6}{(1-z^2)^{-1}\over 57600}+
\\
& &
2(1+z)^{-6}\,\ln\left[(1+z)\,
2^{-1/3}\right],
\\
& &
\\
16M^4\pi ^2D_{{\bar x}{\bar x}{\bar x}{\bar x}}&=&(-3-4z+2z^2)^2
(1+z)^{-8}{(1-z^2)^{-1}\over 24}.\eean

\section{Some useful integrals.}
 
Let us define
\bb  {\cal I}_{n}=\int _{-\infty}^{\infty} {dk}\,\Omega ^n\,
{\rm e}^{-i\epsilon\left(\delta _1\,\Omega -\delta _2\, k\right)}, \;\;\;\;
\Omega =\sqrt{k^2+\lambda};\;\;\; \lambda >0. \label{eq:In}\ee
For the particular value $n=-1$ this integral can be easily solved. Performing 
the following change of
variable,
 
\bb t=t(k)=\delta _1\,\Omega-\delta _2\, k=\delta 
_1\,\sqrt{k^2+\lambda}-\delta _2\, k,    \label{eq:t(k)}\ee
we must consider separately the two possibilities $|\delta _1|>|\delta _2|$ or 
$|\delta _1|<|\delta _2|$.
For $|\delta _1|>|\delta _2|$ and
without loss of generality we may take
$\delta _1,\;\delta _2>0$, and invert the change 
(\ref{eq:t(k)}) as,
 
\bb  k=k(t)=\gamma\,\left[\delta _2\, t\pm\delta _1\,\sqrt{t^2-\beta 
^2}\right],   \label{eq:k(t)}\ee
where we use two new variables
$\gamma$ and $\beta$, defined as,
 
\bb \gamma ^{-1}\equiv \delta _1^2-\delta _2^2,\;\;\; \beta ^2\equiv{\lambda 
\gamma ^{-1}}.              
\label{eq:gam-bet}\ee
Note that $k(t)$, in (\ref{eq:k(t)}), is a bivalued function and we have to be 
careful
in changing the integration limits. Since $|\delta _1|>|\delta _2|$, 
then $\lim _{k\rightarrow\pm\infty} t(k)=+\infty$,
and this means that (\ref{eq:t(k)}) has an absolute minimum at the point
$(+\gamma\,\beta\,\delta _2,\;\beta)$. With this it is easy to see that we 
have to take as inverse function of
(\ref{eq:t(k)}) the function (\ref{eq:k(t)}) with the plus sign to the right 
of the minimum $t=\beta$ and
with the minus sign to the left of the minimum. Therefore we can split the 
integral
(\ref{eq:In}), for $n=-1$, in two parts at each side of $t=\beta$, i.e,
 
\bb{\cal I}_{-1}=\int _{-\infty}^{\gamma\beta\delta _2} {dk\over\Omega}\,
{\rm e}^{-i\epsilon\left(\delta _1\,\Omega -\delta _2\, k\right)}+
\int _{\gamma\beta\delta _2}^{\infty} {dk\over\Omega}\,
{\rm e}^{-i\epsilon\left(\delta _1\,\Omega -\delta _2\, k\right)}. 
\label{eq:I-11}\ee
The change of variables (\ref{eq:t(k)}) gives
$\Omega ^{-1}{dk}=\pm (t^2-\beta ^2)^{-1/2}dt$,
with the minus sign for the first integral in (\ref{eq:I-11}) and the plus 
sign for the second integral. Then
the integration can be easily performed in terms of a zero order Bessel 
function \cite{gra80}, as
 
\bb {\cal I}_{-1}=2\,\int _{\beta}^{\infty} 
{{\rm e}^{-i\epsilon\, t}\over\sqrt{t^2-\beta ^2}}\, dt=
2\,{\rm K}_0 \left(i\epsilon\,\beta\right). \label{eq:I-1R}\ee
 
For the case
$|\delta _1|<|\delta _2|$, we define the parameter $\gamma$, as,
$\gamma ^{-1}\equiv \delta _2^2-\delta _1 ^2$,
and the inverse function of (\ref{eq:t(k)}) is,
 
\bb  k=k(t)=-\gamma\,\left[\delta _2\, t\pm\delta _1\,\sqrt{t^2+\beta 
^2}\right],   \label{eq:k(t)'}\ee
which is again a bivalued function. Now, however, the function
(\ref{eq:t(k)}) has no extrema, and it is easy to see that we can take a 
single inverse everywhere as,
 
$$ k(t)=-\gamma\,\left[\delta _2\, t - \delta _1\,\sqrt{t^2+\beta ^2}\right],
\;\;\;\; {dk\over\Omega}=-{dt\over\sqrt{t^2+\beta ^2}},$$
so that (\ref{eq:In}), for $n=-1$, can also be integrated in terms of a zero 
order Bessel function
\cite{gra80} as,
 
\bb {\cal I}_{-1}=\int _{-\infty}^{\infty} {dk\over\Omega}\,
{\rm e}^{-i\epsilon\left(\delta _1\,\Omega -\delta _2\, k\right)}=
\int _{-\infty}^{\infty} 
{{\rm e}^{-i\epsilon\, t}\over\sqrt{t^2+\beta ^2}}\, dt=
2\,\int _{0}^{\infty} 
{\cos\left(\epsilon\, t\right)\over\sqrt{t^2+\beta ^2}}\, dt
2\,{\rm K}_0 \left(\epsilon\,\beta\right).\label{eq:I-1R'}\ee
 
Finally putting together the results
(\ref{eq:I-1R}) and (\ref{eq:I-1R'}), we can write
 
\bb {\cal I}_{-1}=2\,{\rm K}_0 \left(i\epsilon\,\beta\right),  
\label{eq:I-1RF}\ee
with the parameter $\beta$ given by,
\bb \beta ^2={\lambda \gamma ^{-1}}=\lambda\,\left(\delta _1^2-\delta 
_2^2\right). \label{eq:gam-betR}\ee

The integrals ${\cal I}_{n}$ with $n\geq 0$ are divergent unless we adopt the 
standard prescription
of taking $\epsilon\rightarrow\epsilon -i0^+$ for $|\delta _1|>|\delta _2|$ and
$\epsilon\rightarrow\epsilon +i0^+$ for $|\delta _1|<|\delta _2|$. In that 
case all these integrals
can be easily
related to
${\cal I}_{-1}$ by means of the following recursion formula,
 
\bb {\cal I}_{n}={i\over\epsilon}\,{\partial{\cal I}_{n-1}\over\partial\delta 
_1},\;\;\;\;{\rm then}\;\;\;\;
{\cal I}_{n}=\left({i\over\epsilon}\right)^{n+1}\,
{\partial ^{n+1}{\cal I}_{-1}\over\partial\delta _1^{n+1}}. \label{eq:rec-In}\ee
Since ${\cal I}_{-1}$ is essentially the Bessel function ${\rm K}_0$ and the 
derivatives
of the Bessel functions can be written also in terms of Bessel functions 
\cite{gra80}, then all
${\cal I}_{n}$ with $n\geq 0$ can be expressed as a linear combination of 
Bessel functions.
For example, using (\ref{eq:gam-betR}) for $\beta$ and recalling that
${\rm K}_0'(z)={d{\rm K}_0(z)/ dz}=-{\rm K}_1(z)$,
 
$${\cal I}_{0}={i\over\epsilon}\,{\partial{\cal I}_{-1}\over\partial\delta _1}=
-2\,{\rm K}_0'(i\epsilon\,\beta)\,{\partial\beta\over\partial\delta _1}=
2\,\gamma\,\beta\,\delta _1\, {\rm K}_1(i\epsilon\,\beta) .$$

The integrals ${\cal I}_{n}$ with $n\leq -2 $
are finite in the limit $\epsilon\rightarrow 0$ and 
they  can also be recursively calculated by means of ${\cal I}_{-1}$
and the following recursion relation for $n\leq -1$:
 
\bb {\partial{\cal I}_{n}\over\partial\lambda}={n\over 2}\,{\cal I}_{n-2}-
i{\epsilon\,\delta _1\over 2}\,{\cal I}_{n-1}.              
\label{eq:rec-I-n}\ee
This allows us to calculate all the ${\cal I}_n$ with $n\leq -1$ only if we 
know at least
another integral besides ${\cal I}_{-1}$. Fortunately it is not very difficult 
to evaluate
${\cal I}_{-2}$ from the recursion relation
(\ref{eq:rec-In}) and by a simple integration,
 
$${\cal I}_{-2}(\delta _1)=-i\epsilon\,\int _0^{\delta _1}\, d\delta 
_1'\,{\cal I}_{-1}(\delta _1')+
{\cal I}_{-2}(0), $$
where the integration constant
${\cal I}_{-2}(0)$, which corresponds to the value of 
${\cal I}_{-2}$ for the special value $\delta _1=0$,
can be calculated directly from (\ref{eq:In}) \cite{gra80},
then
 
\bb{\cal I}_{-2}(\delta _1)=-i2\,\epsilon\,\int _0^{\delta _1}\, d\delta _1'\,
{\rm K}_{0}\left(i\epsilon\,\beta (\delta _1')\right)+ 
{\pi \over\lambda ^{1/2}}\, {\rm e}^{\left|\epsilon\,\lambda ^{1/2}\,\delta 
_2\right|}.  \label{eq:I-2}\ee

Let us denote,
\bb  {\cal IE}_{2n}=\int _{-\infty}^{\infty} dk\,\Omega ^{2n}\,
 {\rm Ei}\left(-i\epsilon\,\delta _1\,\Omega\right)
{\rm e}^{i\epsilon\,\delta _2\, k},\;\;\;\;
\Omega =\sqrt{k^2+\lambda};\;\;\; \lambda >0,  \label{eq:IEn}\ee
where ${\rm Ei}(ix)$, with $x$ real, is the integral-exponential function with 
complex
argument \cite{gra80}.  We can easily calculate ${\cal IE}_{0}$ since by the 
properties of ${\rm
Ei}(ix)$, it is not difficult to relate
${\cal IE}_{0}$ with ${\cal I}_{-2}$, 
 
\bb {\partial{\cal IE}_{0}\over\partial\lambda}={1\over 2}\,{\cal I}_{-2}.     
\label{eq:rec-IE0}\ee
This can be easily integrated as
 
\bb {\cal IE}_{0}\left(\lambda\right)={1\over 2}\,\int _0^{\lambda}\, 
d\lambda'\,{\cal I}_{-2}(\lambda')+
{\cal IE}_{0}(0),  \label{eq:IE01}\ee
where, as in the case for
${\cal I}_{-2}$, the arbitrary integration constant ${\cal IE}_0(0)$ can be 
directly calculated from 
(\ref{eq:IEn}) \cite{gra80}, for the particular value $\lambda =0$. The final 
result is,
 
\bb  {\cal IE}_{0}\left(\lambda\right)={1\over 2}\,\int _0^{\lambda}\, 
d\lambda'\,{\cal I}_{-2}(\lambda')+
\left\{\begin{array}{cl} 
\displaystyle {i\over \epsilon\,\delta _2}\,\ln\left[(\delta _1+\delta _2)^2\, 
\gamma\right],&
\;\;\;{\rm for}\;\;  \delta _2\neq 0.\\ 
\displaystyle -{2i\over \epsilon\,\delta _1},&
\;\;\;{\rm for}\;\;  \delta _2= 0. 
\end{array}\right.    \label{eq:IE0RF}\ee
 
The integrals ${\cal IE}_{2\, n}$ with $n>0$ diverge unless we adopt the 
standard prescription
$\epsilon\rightarrow\epsilon -i0^+$ for $|\delta _1|>|\delta _2|$ and
$\epsilon\rightarrow\epsilon +i0^+$ for $|\delta _1|<|\delta _2|$ then they
can be calculated with
the  known values of
${\cal I}_n$ and ${\cal IE}_0$ by means of the following expression,
 
$$ {\partial{\cal I}_{n}\over\partial\delta _2}=i\epsilon\,
\int _{-\infty}^{\infty} dk\, \left({{\rm e}^{-i\epsilon\,\delta 
_1\,\Omega}\over\Omega}\,
{k\over\Omega}\right)\,\Omega ^{n+2}\,
{\rm e}^{i\epsilon\,\delta _2\, k}, $$
which can be integrated by parts, using the properties of the ${\rm Ei}(ix)$, 
to obtain the following recursion
relation
 
\bb {\cal IE}_{n+2}={1\over \epsilon ^2\,\delta 
_2}\,{\partial\over\partial\delta _2}\,
\left[{\cal I}_{n}+(n+2)\,{\cal IE}_{0}\right]. \label{eq:rec-IEn}\ee

\section{Geodesic coefficients $\xi _i^*$ and $x_i$}

We start with either a timelike or spacelike geodesic connecting the points 
$x'$, $\bar x$ and $x$, which
can be written in terms of the proper geodesic distance $\tau$ as,
 
$$x^{\mu}=x^{\mu}\left({\bar\tau}+\epsilon\right),\;\;\;\;{\bar
x}^{\mu}=x^{\mu}\left({\bar\tau}\right),\;\;\;\;
x'^{\mu}=x^{\mu}\left({\bar\tau}-\epsilon\right),  $$ 
and define at the midpoint
$\bar x$ the parameters
$\xi _n^*=\left.{d^n\xi ^* / d\tau ^n}\right|_{\bar\tau}$ and $x_n=\left.{d^nx 
/ d\tau
^n}\right|_{\bar\tau}$, which determine the geodesic. From the metric in the 
interaction region keeping 
$\eta$ and $y$ fixed, by the simmitry of the problem, we get
 
\bb ds^2=\left({1-\sin\xi\over 1+\sin\xi}\right)\,\left(d\xi ^{*2}-dx^2\right) 
. \label{eq:m1}\ee
If we
parametrize the geodesic with its proper distance, then 
(\ref{eq:m1}) gives the following relation between the coefficients $\xi _1^*$ 
and $x_1$, 
 
\bb  \xi _1^{*2}-x_1^2=\pm {1+\sin{\bar\xi}\over 1-\sin{\bar\xi}} ,  
\label{eq:xi1x1}\ee
with the plus sign for timelike geodesics and the minus sign for spacelike 
geodesics.
Since the coordinate $x$ does not appear in (\ref{eq:m1}), it is easy to see 
that,
 
\bb {dx\over d\tau}=-p_x\,\left({1+\sin{\xi}\over 1-\sin{\xi}}\right),\;\;\;\;
 \left({d\xi ^*\over d\tau}\right)^2=p_x^2\,\left({1+\sin{\xi}\over 
1-\sin{\xi}}\right)^2 \pm {1+\sin{\xi}\over
1-\sin{\xi}}.\label{eq:dxdxi}\ee
Equations (\ref{eq:dxdxi}) and the relation between $\xi$ and $\xi ^*$,
$d\xi ^*=d\xi\, M\,{(1+\sin\xi)^2/\cos\xi}$,
allow us to obtain the derivatives
$d^n\xi ^*/d\tau ^n$ and $d^n x /d\tau ^n$, which when evaluated at
$\tau ={\bar\tau}$ produce the geodesic coefficients we need.

\end{document}